\documentclass[12pt]{article}
\usepackage{epsfig,cite}
\usepackage {amsmath}
\usepackage{amssymb}
\include{epsf}
\newcount\timecount
\newcount\hours \newcount\minutes  \newcount\temp \newcount\pmhours
\hours = \time
\divide\hours by 60
\temp = \hours
\multiply\temp by 60
\minutes = \time
\advance\minutes by -\temp
\def\hour{\the\hours}
\def\minute{\ifnum\minutes<10 0\the\minutes
            \else\the\minutes\fi}
\def\clock{
\ifnum\hours=0 12:\minute\ AM
\else\ifnum\hours<12 \hour:\minute\ AM
      \else\ifnum\hours=12 12:\minute\ PM
            \else\ifnum\hours>12
                 \pmhours=\hours
                 \advance\pmhours by -12
                 \the\pmhours:\minute\ PM
                 \fi
            \fi
      \fi
\fi
}

\def\monthname{\relax\ifcase\month 0/\or January\or February\or
   March\or April\or May\or June\or July\or August\or September\or
   October\or November\or December\else\number\month/\fi}

\def\bold#1{\setbox0=\hbox{$#1$}%
     \kern-.025em\copy0\kern-\wd0
     \kern.05em\copy0\kern-\wd0
     \kern-.025em\raise.0433em\box0 }

\def\beq{\begin{equation}}
\def\eeq{\end{equation}}

\def\gev{{\rm \, Ge\kern-0.125em V}}
\def\tev{{\rm \, Te\kern-0.125em V}}
\def\gyr{{\rm \, G\kern-0.125em yr}}
\def\ohsq{\Omega_{\chi} h^2}

\def\nl{\hfill\nonumber\\&&}
\def\nnl{\hfill\nonumber\\}

\def\Toprel#1\over#2{\mathrel{\mathop{#2}\limits^{#1}}}

\def\m12{m_{1\!/2}}

\def\bea{\begin{eqnarray}}
\def\eea{\end{eqnarray}}

\def\mplr{\overline{M_{P}}}
\def\mgut{M_{\rm GUT}}

\def\afiv{A_\mathbf{\overline{5}}}
\def\aten{A_\mathbf{10}}

\mathchardef\mhyphen="2D

\DeclareMathOperator{\Tr}{Tr}

\def\mplr{\overline{M_{P}}}
\def\yfiv{y_\mathbf{\overline{5}}}
\def\yten{y_\mathbf{10}}
\def\yone{y_\mathbf{1}}
\def\bfyfiv{\mathbf{\yfiv}}
\def\bfyten{\mathbf{\yten}}
\def\bfyone{\mathbf{\yone}}
\def\bfys{\mathbf{y}_S}
\def\afiv{A_\mathbf{\overline{5}}}
\def\aten{A_\mathbf{10}}
\def\aone{A_\mathbf{1}}
\def\alfor{A_{\lambda_4}}
\def\alfiv{A_{\lambda_5}}

\begin{document}
\begin{titlepage}
\pagestyle{empty}
\baselineskip=21pt
\rightline{CERN-PH-TH/2011-037, KCL-PH-TH/2011-07}
\rightline{UMN--TH--2940/11, FTPI--MINN--11/07}
\rightline{SU--ITP--11/18, SLAC--PUB--14411}
\vskip 0.2in
\begin{center}
{\large{\bf Constrained Supersymmetric Flipped SU(5) GUT Phenomenology}}
\end{center}
\begin{center}
\vskip 0.2in
{\bf John~Ellis}$^1$, {\bf Azar Mustafayev}$^{2}$ and {\bf Keith~A.~Olive}$^{2,3}$

\vskip 0.1in

{\it
$^1${TH Division, PH Department, CERN, CH-1211 Geneva 23, Switzerland, \\
Theoretical Physics and Cosmology Group, Department of Physics, King's College London, London WC2R 2LS, UK}\\
$^2${William I.~Fine Theoretical Physics Institute, \\
University of Minnesota, Minneapolis, MN 55455, USA}\\
$^3${Department of Physics and SLAC,
Stanford University, Palo Alto, CA 94305}\\
}

\vskip 0.2in
{\bf Abstract}
\end{center}
\baselineskip=18pt \noindent

We explore the phenomenology of the minimal supersymmetric flipped SU(5) GUT model (CFSU(5)),
whose soft supersymmetry-breaking (SSB) mass parameters are constrained to be
universal at some  input scale, $M_{in}$, above the GUT scale, $\mgut$. We analyze
the parameter space of CFSU(5) assuming that the lightest supersymmetric
particle (LSP) provides the cosmological cold dark matter, paying careful
attention to the matching of parameters at the GUT scale.
We first display some specific examples of the evolutions of the SSB parameters that exhibit some generic features. 
Specifically, we note that the relationship between the masses of the lightest neutralino $\chi$ and the lighter 
stau ${\tilde \tau_1}$ is sensitive to $M_{in}$, as is the relationship between $m_\chi$ and the masses of the
heavier Higgs bosons $A, H$. For these reasons, prominent features in generic $(m_{1/2}, m_0)$ planes
such as coannihilation strips and rapid-annihilation funnels are also sensitive to $M_{in}$, as we
illustrate for several cases with $\tan \beta = 10$ and 55. However, these features do not necessarily disappear
at large $M_{in}$, unlike the case in the minimal conventional SU(5) GUT. Our results are relatively insensitive
to neutrino masses.

\vspace{1cm}
\vfill
\leftline{CERN-PH-TH/2011-037}
\leftline{March 2011}
\end{titlepage}

\section{Introduction}

The principal bugbear in supersymmetric phenomenology is our ignorance of the mechanism
for supersymmetry breaking and hence its effective pattern at low energies. The observed 
suppression of flavour-changing neutral interactions motivates universality for the
soft supersymmetry-breaking (SSB) scalar mass parameters for different sfermions
with the same Standard Model quantum numbers~\cite{EN,BG}, and
Grand Unified Theories (GUTs) suggest universality between the SSB scalar masses of squarks and sleptons in
the same GUT multiplets. There have been many studies of the model in which the SSB scalar
masses of all squarks, sleptons and Higgs multiplets are constrained to be universal at
some input scale, $M_{in}$ usually taken to be the GUT scale, $\mgut$, 
(the CMSSM)~\cite{funnel,cmssm,efgosi,cmssmnew,cmssmmap}.

However, it could be argued that a more natural choice of scale for universality would be some scale
associated with supergravity or string compactification, above the GUT scale. One specific
example that has been studied is the minimal supersymmetric SU(5) GUT~\cite{EMO}, where it has
been shown that the resulting phenomenology is quite sensitive to the choice of $M_{in}$,
and potentially very different from the conventional CMSSM case. However, the 
low-energy phenomenology will, in general, depend on the choice of GUT group. This is because, in
particular, the running of the SSB parameters between $\mgut$ and $M_{in}$ depends on the 
choice of GUT gauge group, and the choice of universality conditions on the SSB parameters at 
$M_{in}$ depends on the GUT multiplet assignments.

With this motivation,
in this paper we study the phenomenological property of another GUT model, namely minimal 
supersymmetric flipped SU(5)~\footnote{This minimal scenario is different from the variant of flipped SU(5) 
derived from F-theory, $\mathcal{F}$-SU(5), proposed and studied in~\cite{Nanopoulos}, which also includes 
vector-like particles below the GUT scale.}.
This GUT has several advantages over the SU(5) GUT -- the doublet-triplet splitting problem is 
resolved with use of only minimal Higgs
representations and protons are naturally long lived~\cite{Antoniadis:1989zy}, 
neutrinos are necessarily massive~\cite{Ellis:1992nq,Leontaris:1992wp}, and 
supersymmetric hybrid inflation can easily be implemented successfully~\cite{infl}.
On the other hand, these advantages come at the expense of some clear disadvantages: the successful conventional 
supersymmetric SU(5) prediction for the weak mixing angle $\sin^2 \theta_W$ is lost, as is the corresponding
prediction for $m_b$ based on Yukawa unification, and the model has additional parameters, as discussed below.

We assume in our analysis of flipped SU(5) 
that the SSB parameters are universal at some high input scale $M_{in}$, a framework
we term constrained flipped SU(5), or CFSU(5).
We explore the sensitivity of CFSU(5) phenomenology to the choice of $M_{in}$,
to the choice of Yukawa couplings in the model and to the range of neutrino masses, 
and contrast our findings with those in the CMSSM and the minimal SU(5) GUT. 

As is well known in the case of the CMSSM, there are coannihilation strips 
and rapid-annihilation funnels compatible~\cite{cmssmmap} with estimates 
of the cosmological cold dark matter density based on WMAP and other data~\cite{WMAP}. 
We find that these move significantly in the $(m_{1/2}, m_0)$ planes of CFSU(5) as $M_{in}$ is varied in the
range up to $2.4 \times 10^{18}$~GeV. However, unlike the case of minimal conventional SU(5)
studied in~\cite{EMO}, these WMAP-compatible regions do not disappear entirely as $M_{in}$ increases.
These WMAP-compatible regions are also sensitive to unconstrained Yukawa couplings in the CFSU(5)
model, and to a lesser extent to the neutrino mass scale. In order to accentuate the effects of the neutrino sector,
we consider the case of a `large' neutrino mass $\sim 0.3$~eV, but only in exceptional cases do we find any
significant difference between this and the more conservative choice $\sim 0.05$~eV.

The structure of this paper is as follows. In Section~\ref{sec:fsu5} we specify the parameters of the CFSU(5)
model, and recapitulate the RGEs for its supersymmetric couplings and soft supersymmetry-breaking parameters.
In Section~\ref{sec:RGEs} we explore numerically the effects of the renormalizations of these parameters, exploring
in particular the relation between the LSP $\chi$ and the ${\tilde \tau}_1$ mass - which affects the location
of the coannihilation strip - and that between the $\chi$ and the heavy Higgs bosons $A/H$ - which affects
the location of the rapid-annihilation funnel. Section~\ref{sec:planes} presents some generic $(m_{1/2}, m_0)$
planes, and uses them to discuss the influences of $M_{in}$, Yukawa couplings and the neutrino mass. 
Finally, Section~\ref{sec:concl} summarizes our conclusions.

\section{The Minimal Flipped SU(5) GUT Superpotential and RGEs}
\label{sec:fsu5}

The minimal flipped SU(5) GUT model~\cite{fsu5,Antoniadis:1987dx} (denoted hereafter
by FSU(5)) is based on the gauge group SU(5) $\times$ U(1)$_X$, which is a maximal subgroup of SO(10). 
Defining the hypercharge generator in SU(5) as 
\beq
T_{24}=\sqrt{\frac{3}{5}}diag\left( \frac{1}{3},\frac{1}{3},\frac{1}{3},-\frac{1}{2},-\frac{1}{2}\right) ,
\eeq
the Standard Model hypercharge $Y$ is the following linear combination of SU(5) and U(1)$_X$ charges:
\beq
Y/2 = (-Y_5/2 +\sqrt{24}X)/5 \, .
\eeq
The matter sector of minimal FSU(5) contains three families of chiral superfields:
\bea
\hat{f}_i(\mathbf{\overline{5}},-3)=\{\hat{U}^c_i, \hat{L}_i\} \,,& 
\hat{F}_i(\mathbf{10},1)=\{\hat{Q}_i, \hat{D}^c_i, \hat{N}^c_i\} \,, &
\hat{l}_i(\mathbf{1},5)= \hat{E}^c_i \,,
\eea
where $i= 1, 2, 3$ is a generation index, and the numbers in parentheses denote transformation
properties under the SU(5) $\times$ U(1)$_X$ gauge group, with the U(1)$_X$ charges expressed in units of
$1/\sqrt{40}$. In addition, to generate heavy right-handed neutrino masses, we introduce three FSU(5) singlets 
$\hat{S}_i(\mathbf{1},0)$. 
The MSSM electroweak Higgs doublets $\hat{H}_u$ and $\hat{H}_d$ reside in five-dimensional 
representations
\bea
\hat{h}_1(\mathbf{5},-2)=\{\hat{T}_1, \hat{H}_d\} \,,&&
\hat{h}_2(\mathbf{\overline{5}},2)=\{\hat{T}_2, \hat{H}_u\} \, .
\eea
and we also introduce a pair of GUT Higgs multiplets
\bea
\hat{H}_1(\mathbf{10},1)=\{\hat{Q}_{H_1}, \hat{D}^c_{H_1}, \hat{N}^c_{H_1}\}\,,&&
\hat{H}_2(\mathbf{\overline{10}},-1)=\{\hat{Q}_{H_2}, \hat{D}^c_{H_2}, \hat{N}^c_{H_2}\} \, .
\eea
to break FSU(5) down to the Standard Model gauge group. 

The minimal renormalizable superpotential is
\bea
W &=& (\bfyfiv)_{ij}\hat{F}_i^{\alpha\beta}\hat{h}_{2\alpha}\hat{f}_{j\beta}
     -(\bfyten)_{ij}\epsilon_{\alpha\beta\gamma\delta\zeta}
       \hat{F}_i^{\alpha\beta}\hat{F}_j^{\gamma\delta}\hat{h}_1^\zeta
     -(\bfyone)_{ij}\hat{f}_{i\alpha}\hat{l}_{j}\hat{h}_1^\alpha
     +\mu_h\hat{h}_1^\alpha\hat{h}_{2\alpha} \nl
     +\lambda_4\epsilon_{\alpha\beta\gamma\delta\zeta}
      \hat{H}_1^{\alpha\beta}\hat{H}_1^{\gamma\delta}\hat{h}_1^\zeta
     +\lambda_5\epsilon^{\alpha\beta\gamma\delta\zeta}
     \hat{H}_{2\alpha\beta}\hat{H}_{2\gamma\delta}\hat{h}_{2\zeta}
     +(\bfys)_{ij}\hat{F}_i^{\alpha\beta}\hat{H}_{2\alpha\beta}\hat{S}_j
     +(\mu_S)_{ij}\hat{S}_i\hat{S}_j ,
\label{Wfsu5}
\eea
where Greek letters denote SU(5) indices and $\epsilon$ is
the totally antisymmetric tensor with $\epsilon_{12345}=\epsilon^{12345}=1$.
We assume the discrete symmetry $\hat{H}_1\rightarrow -\hat{H}_1$, which prevents the mixing of ordinary
fermions with color triplets $\hat{T}_1$ and members of Higgs decuplets through couplings 
$\hat{F}\hat{H}_1\hat{h}_1$ and $\hat{H}_1\hat{f}\hat{h}_2$.
We recall that large vevs of the Higgs decuplets $\langle N^c_{H_1}\rangle =\langle N^c_{H_2}\rangle =V$
generate couplings between the colour triplets $\hat{T}_1$ and $\hat{T}_2$ with $\hat{D}^c_{H_1}$ and
$\hat{D}^c_{H_2}$, respectively, forming heavy states with
masses $\lambda _4 V$ and $\lambda _5 V$. Note that any domain walls
formed during the breaking of the discrete symmetry when the decuplets obtain vevs
are expected to be inflated away and are thus harmless so long as the
reheating scale after inflation is below $V$.  In addition, doublet-triplet splitting occurs via a very economical
missing-partner mechanism, one of the attractive features of FSU(5).

The pentaplet mixing term $\mu_h\hat{h}_1\hat{h}_2$ could arise from a small vev of an FSU(5) singlet 
field, {\it i.e.}, $\lambda_7\hat{h}_1\hat{h}_2\phi \rightarrow \lambda_7
\langle\phi\rangle\hat{h}_1\hat{h}_2$~\cite{Antoniadis:1987dx,Antoniadis:1989zy},
or from an effective higher-order coupling~\cite{mu2}, or from a supergravity mechanism~\cite{mu3}.
The coupling $\mu_h$ should be small in order to avoid rapid dimension-five proton decay arising 
from the exchange of colored higgsinos $\tilde{T}_{1,2}$.
Here, we take a phenomenological approach, simply assuming that $\mu_h$ is of the order of the electroweak scale.  
A similar argument can be made for the presence of $(\mu_S)_{ij} \gg y_S V$ in Eq.~\ref{Wfsu5}.  
Furthermore, $R$-parity would prevent additional couplings such 
$(\lambda^{'}_7)_i \hat{h}_1\hat{h}_2 S_i$, that would unnecessarily complicate 
the calculation of the dark matter relic density in this model.

Another attractive feature of FSU(5) is that it naturally contains singlet (right-handed) 
neutrinos~\cite{Ellis:1992nq,Leontaris:1992wp}.
This allows for the generation of small neutrino masses through the 
mechanism known as the double 
seesaw~\cite{dseesaw}, which utilizes the seesaw formula twice. 
The neutrino mass matrix receives contributions from the $F h_2 f,\ F H_2 S$ and $SS$ terms in the
superpotential~(\ref{Wfsu5}). The resulting $9\times 9$ matrix, in the $(\nu_i, N^c_i, S_i)$ basis, has the form
\beq
\label{eq:seesaw}
\mathcal{M}_{\nu}=
 \begin{pmatrix}
 0 & \mathbf{h}_\nu v_u& 0 \\
 \mathbf{h}^T_\nu v_u& 0 & \bfys V\\
 0 &\bfys^T V & \mu_S 
 \end{pmatrix} .
\eeq
In the first stage, the singlets $S_i$ decouple, generating Majorana masses for the right-handed neutrino fields $N^c_i$,
$M_N\simeq V^2 \bfys \mu_S^{-1} \bfys^T$. 
In the second stage, decoupling of the $N^c_i$ generates the desired small masses for the left-handed neutrinos,
$m_\nu \simeq v^2_u \mathbf{h}_\nu M_N^{-1} \mathbf{h}^T_\nu$. 

In discussing the couplings and renormalization-group equations (RGEs) for FSU(5), we assume third-generation
dominance, {\it i.e.}, we neglect the Yukawa couplings of the first two generations,
so that ${\bf y_{{\overline{5}},10,1,S}} \sim \left({\bf y_{{\overline{5}},10,1,S}}\right)_{33} 
\equiv y_{\mathbf{\overline{5}},\mathbf{10},\mathbf{1},S}$.
In this scheme the mass of the $\tau$ neutrino is
\beq
m_{\nu_\tau}\simeq\frac{v_u^2h_\nu^2 \mu_S}{y_S^2 V^2} .
\eeq

We match the gauge and superpotential couplings of FSU(5) to those of the MSSM at the scale
where its SU(2)$_L$ and SU(3)$_c$ gauge couplings are unified, denoted hereafter by $\mgut$:
\bea
 \alpha_2=\alpha_3 = \alpha_5 &&
 25\alpha_1^{-1} = 24\alpha_X^{-1}+\alpha_5^{-1}  \nnl
 h_t = h_{\nu} = \yfiv /\sqrt{2}&&
 h_b = 4\yten  \nnl
 h_\tau = \yone && \mu = \mu_h
\label{matching1}
\eea
where $\alpha_1\equiv 5/3 g_Y^2/(4\pi)$.
Note that unlike minimal $SU(5)$, here the neutrino Yukawa coupling is naturally fixed to be equal to the up-quark
Yukawa coupling. This is a consequence of the flipping that puts the right-handed neutrinos (RHNs)  
into decuplets in FSU(5), instead of being singlets as
in minimal SU(5), where the Yukawa coupling would be viewed as an independent parameter. 

The following are the SSB terms entering our analysis:
\bea
\mathcal{L}_{soft} &=& -m_{h_1}^2 h_{1}^\dagger h_1 -m_{h_2}^2 h_2^\dagger h_{2}
          -m_{H_1}^2 \Tr\{H_1^\dagger H_1\} -m_{H_2}^2 \Tr\{H_2^\dagger H_2\} \nl
          -(\mathbf{m_f^2})_{ij} \tilde{f}_i^\dagger \tilde{f}_j
          -(\mathbf{m_F^2})_{ij} \Tr\{\tilde{F}_i^\dagger \tilde{F}_j\}
          -(\mathbf{m_l^2})_{ij} \tilde{l}_i^\dagger \tilde{l}_j
          -(\mathbf{m_S^2})_{ij} \tilde{S}_i^\dagger \tilde{S}_j \nl
	  -\frac{1}{2}M_5 \overline{\tilde{\lambda}_5}\tilde{\lambda}_5
	  -\frac{1}{2}M_X \overline{\tilde{\lambda}_X}\tilde{\lambda}_X \nl
	  -\left[ \afiv\yfiv \tilde{F}_i h_2\tilde{f}_j  
	          -\aten\yten \tilde{F}_i \tilde{F}_j h_1 
		  -\aone\yone \tilde{f}_i \tilde{l}_j h_1
	          +\alfor\lambda_4 H_1 H_1 h_1 +\alfiv\lambda_5 H_2 H_2 h_2 \right.\nl
           \left.\quad+A_S y_S \tilde{F}_i H_2 \tilde{S}_j 
	   +B_h \mu_h h_1 h_2  + B_S \mu_S \tilde{S}_i \tilde{S}_j +\mathrm{h.c.}
	   \right]_{i=j=3}\, ,
\label{SSBfsu5}
\eea
where we have suppressed the SU(5) indices. In order to comply with 
stringent flavor-violating constraints~\cite{EN,Ciuchini:2007ha}, 
we assume that SSB scalar mass matrices are flavor-diagonal and have degenerate first- and second-generation entries. 
For example, we assume the decuplet sfermion mass matrix to have the form 
$\mathbf{m_F^2}=diag(m_{F_1}^2,m_{F_1}^2,m_F^2)$,
with similar expressions for 
$\mathbf{m_f^2},\ \mathbf{m_l^2}$ and $\mathbf{m_S^2}$ SSB matrices.

The matching conditions for the SSB terms at $\mgut$ are
\bea
 M_2=M_3 = M_5 &&
 25M_1\alpha_1^{-1} = 24M_X\alpha_X^{-1}+M_5\alpha_5^{-1} \nnl
 m_{Q_1}^2=m_{D_1}^2=m_{N_1}^2 = m_{F_1}^2 &&
 m_{Q_3}^2=m_{D_3}^2=m_{N_3}^2 = m_F^2 \nnl
 m_{U_1}^2=m_{L_1}^2 = m_{f_1}^2 &&
 m_{U_3}^2=m_{L_3}^2 = m_f^2 \nnl
 m_{E_1}^2 = m_{l_1}^2 &&
 m_{E_3}^2 = m_l^2 \nnl
 m_{H_u}^2 = m_{h_2}^2 &&
 m_{H_d}^2 = m_{h_1}^2 \nnl
 A_t=A_\nu = \afiv &&
 A_b = \aten \nnl
 A_\tau = \aone &&  B = B_h .
\label{matching2}
\eea
We assume in our phenomenological analysis universality of the FSU(5) SSB terms at the scale $M_{in}$,
which provides us with an additional set of boundary conditions at $M_{in}$:
\bea
m_{f_1}=m_{F_1}=m_{l_1}=m_f=m_F=m_l=m_{h_1}=m_{h_2}=m_{H_1}=m_{H_2} = m_S &\equiv& m_0 , \nnl
M_X = M_5 &\equiv& m_{1/2} , \nnl
\afiv=\aten=\aone=\alfor=\alfiv =A_S &\equiv& A_0 ,
\eea
and defines the scenario we term constrained FSU(5), or CFSU(5).
This model is completely specified by the following set of parameters:
\beq
m_0,\ m_{1/2},\ A_0,\ \tan\beta,\ M_{in},\ \lambda_4(\mgut),\ \lambda_5(\mgut),\ y_S(\mgut),\ sgn(\mu),\ m_t,\ m_{\nu_3} .
\label{CFSU5params}
\eeq

In this work we set the mass of the top quark $m_t = 173.1$~GeV in accordance with the latest Tevatron results~\cite{mt},
set the running bottom quark mass $m_b^{\overline{MS}}(m_b) = 4.2$~GeV~\cite{rpp},  
and choose $m_{\nu_3} = 0.3$~eV or 0.05~eV. 
We consider $sgn(\mu)=+1$, which is favored by $g_\mu - 2$~\cite{newBNL,g-2} and $BR(b \to s \gamma)$~\cite{bsgex} measurements. 
We employed the program {\tt SSARD}~\cite{ssard} to perform RGE evolutions at 2-loop level for the MSSM and at 1-loop level for
CFSU(5). The matching between the two theories is done at the scale $\mgut$ according to expressions
(\ref{matching1}, \ref{matching2}). The location of $\mgut$ is determined dynamically as the scale where the two non-abelian MSSM gauge
couplings meet, $g_2=g_3$. This definition of $\mgut$ (also called $M_{23}$) is somewhat different 
from the usual definition, according to which $\mgut$ is the scale where $g_1=g_2$. Those two scales are very close to each other, so we use $\mgut$ 
as the scale up to which the SM gauge group is valid. 
We do not require unification of Abelian and non-Abelian gauge couplings, 
although this might occur for choices of $M_{in}$ and would be motivated by some string scenarios.
The weak-scale RGE parameters are further passed to the {\tt FeynHiggs~2.6.5} 
code~\cite{FeynHiggs} for computation of the light CP-even Higgs boson mass $m_h$.
We also performed cross-checks using the {\tt ISAJET 7.80}~\cite{isajet} program, 
augmented with FSU(5) evolution, and found results
in good agreement~\footnote{For a comparison of the {\tt SSARD} and {\tt ISAJET} codes, see Ref.~\cite{Battaglia:2001zp}.}.

Between $M_{in}$ and $\mgut$, the applicable one-loop RGEs for the FSU(5) gauge couplings are:
\bea
\frac{d g_5^2}{dt} & = & \frac{g_5^4}{8\pi^2} (-5) \, , \\
\frac{d g_X^2}{dt} & = & \frac{g_X^4}{8\pi^2}\frac{15}{2}  \, ,
\label{RGEgauge}
\eea
and the applicable one-loop RGEs for the Yukawa couplings in the superpotential (\ref{Wfsu5}) are:
\bea
\frac{d \yfiv}{dt} & = & 
  \frac{\yfiv}{16 \pi^2}\left[ 48\yten^2 + 5\yfiv^2 + \yone^2 + y_S^2 + 48\lambda_5^2
    -\frac{84}{5}g_5^2 -\frac{7}{10}g_X^2 \right], 
    \eea
    \bea
\frac{d \yten}{dt} & = & 
  \frac{\yten}{16 \pi^2}\left[ 144\yten^2 + 2\yfiv^2 +\yone^2 + 2 y_S^2 + 48\lambda_4^2
    -\frac{96}{5}g_5^2 -\frac{3}{10}g_X^2 \right], 
      \eea
    \bea
\frac{d \yone}{dt} & = & 
  \frac{\yone}{16 \pi^2}\left[ 48\yten^2 + 2\yfiv^2 + 7\yone^2 + 48\lambda_4^2
    -\frac{48}{5}g_5^2 -\frac{19}{10}g_X^2 \right], 
      \eea
    \bea
\frac{d \lambda_4}{dt} & = & 
  \frac{\lambda_4}{16 \pi^2}\left[ 48\yten^2 + \yone^2 + 144\lambda_4^2
    -\frac{96}{5}g_5^2 -\frac{3}{10}g_X^2 \right], \label{RGElam4}
      \eea
    \bea
\frac{d \lambda_5}{dt} & = & 
  \frac{\lambda_5}{16 \pi^2}\left[ 2\yfiv^2 + 2y_S^2 + 144\lambda_5^2
    -\frac{96}{5}g_5^2 -\frac{3}{10}g_X^2 \right], \label{RGElam5}
      \eea
    \bea
\frac{d y_S}{dt} & = & 
  \frac{y_S}{16 \pi^2}\left[ 48\yten^2 + \yfiv^2 + 12y_S^2 + 48\lambda_5^2
    -\frac{72}{5}g_5^2 -\frac{1}{10}g_X^2 \right].
\label{RGEYukawa}
\eea
As usual, the coefficients in the one-loop RGEs for the SSB gaugino masses are the same as those
in the RGEs for the gauge couplings squared. 
The one-loop RGEs for the third-generation SSB scalar masses are:
\bea
\frac{d m_{h_1}^2}{dt} & = &
  \frac{1}{16 \pi^2}\left[ 96\yten^2 \left( m_{h_1}^2 +2m_F^2 +\aten^2 \right)
    +2\yone^2 \left( m_f^2 +m_l^2 +m_{h_1}^2 +\aone^2 \right) \right.\nl
 \left. \quad\qquad +96\lambda_4^2 \left( m_{h_1}^2 +2m_{H_1}^2 +\alfor^2 \right)
    -g_X^2 \mathcal{S}
    -\frac{96}{5} g_5^2 M_5^2 -\frac{4}{5} g_X^2 M_X^2 \right] \,,   \label{RGEmass0} \\
\frac{d m_{h_2}^2}{dt} & = &
  \frac{1}{16 \pi^2}\left[ 4\yfiv^2 \left( m_{h_2}^2 +m_F^2 +m_f^2 +\afiv^2 \right)
    +96\lambda_5^2 \left( m_{h_2}^2 +2m_{H_2}^2 +\alfiv^2 \right) \right.\nl
 \left. \quad\qquad 
    +g_X^2 \mathcal{S}
    -\frac{96}{5} g_5^2 M_5^2 -\frac{4}{5} g_X^2 M_X^2 \right] \,, \\
\frac{d m_{H_1}^2}{dt} & = &
  \frac{1}{16 \pi^2}\left[ 96\lambda_4^2 \left( m_{h_1}^2 +2m_{H_1}^2 +\alfor^2 \right)
    +\frac{1}{2} g_X^2 \mathcal{S}
    -\frac{144}{5} g_5^2 M_5^2 -\frac{1}{5} g_X^2 M_X^2 \right] \,, \\
\frac{d m_{H_2}^2}{dt} & = &
  \frac{1}{16 \pi^2}\left[ 96\lambda_5^2 \left( m_{h_2}^2 +2m_{H_2}^2 +\alfiv^2 \right)
    +2y_S^2 \left( m_F^2 +m_{H_2}^2 +m_S^2 +A_S^2 \right) \right.\nl
 \left. \quad\qquad 
    -\frac{1}{2} g_X^2 \mathcal{S}
    -\frac{144}{5} g_5^2 M_5^2 -\frac{1}{5} g_X^2 M_X^2 \right] \,, \\
\frac{d m_f^2}{dt} & = &
  \frac{1}{16 \pi^2}\left[ 4\yfiv^2 \left( m_{h_2}^2 +m_F^2 +m_f^2 +\afiv^2 \right)
      +2\yone^2 \left( m_f^2 +m_l^2 +m_{h_1}^2 +\aone^2 \right) \right.\nl
 \left. \quad\qquad 
    -\frac{3}{2}g_X^2 \mathcal{S}
    -\frac{96}{5} g_5^2 M_5^2 -\frac{9}{5} g_X^2 M_X^2 \right] \,, \\
\frac{d m_F^2}{dt} & = &
  \frac{1}{16 \pi^2}\left[ 96\yten^2 \left( m_{h_1}^2 +2m_F^2 +\aten^2 \right)
    +2\yfiv^2 \left( m_f^2 +m_{h_2}^2 +m_F^2 +\afiv^2 \right) \right.\nl
 \left. \quad\qquad +2y_S^2 \left( m_F^2 +m_S^2 +m_{H_2}^2 +A_S^2 \right) 
    +\frac{1}{2}g_X^2 \mathcal{S}
    -\frac{144}{5} g_5^2 M_5^2 -\frac{1}{5} g_X^2 M_X^2 \right] \,, \\
\frac{d m_l^2}{dt} & = &
  \frac{1}{16 \pi^2}\left[ 10\yone^2 \left( m_f^2 +m_l^2 +m_{h_1}^2 +\aone^2 \right)
    +\frac{5}{2}g_X^2 \mathcal{S}
    -5 g_X^2 M_X^2 \right] \,, \label{RGEmassl}\\
\frac{d m_S^2}{dt} & = &
  \frac{1}{16 \pi^2} 20y_S^2 \left( m_F^2 +m_{H_2}^2 +m_S^2 +A_S^2 \right) \,,
\label{RGEmass}
\eea
where
\beq
\mathcal{S}\equiv \Tr\{ Xm^2\}= m_{H_1}^2 -m_{H_2}^2 -m_{h_1}^2 +m_{h_2}^2
   +\Tr \left( \mathbf{m_F^2}-\frac{3}{2}\mathbf{m_f^2}+\frac{1}{2}\mathbf{m_l^2} \right) \, .
\eeq
is the analog of the S-term in MSSM RGEs~\cite{MSSMRGEs}. Note that in a universal scenario $\mathcal{S}=0$ at $M_{in}$  
and remains zero at all scales, as expected for an anomaly-free theory. 
The RGEs for the SSB mass-squared parameters of the first and second generations,
$m_{f_1}^2,\ m_{F_1}^2$, $m_{l_1}^2,\ m_{S_1}^2$, can be obtained from the above
RGEs for their third-generation counterparts, $m_f^2,\ m_F^2$, $m_l^2,\ m_S^2$, 
simply by removing the terms involving the Yukawa couplings
$\yten, \yfiv, \yone, y_S$ that give masses to the third-generation fermions.
The one-loop RGEs for the SSB trilinear $A$ terms are:
\bea
\frac{d \afiv}{dt} & = & 
  \frac{1}{16 \pi^2}\left[ 96\aten\yten^2 + 10\afiv\yfiv^2 + 2\aone\yone^2 + 2A_S y_S^2
    +96\alfiv\lambda_5^2   \right.\nl
    \left.\quad\qquad
    -\frac{168}{5} g_5^2 M_5 -\frac{7}{5} g_X^2 M_X \right], \\
\frac{d \aten}{dt} & = & 
  \frac{1}{16 \pi^2}\left[ 288\aten\yten^2 + 4\afiv\yfiv^2 + 2\aone\yone^2 + 4A_S y_S^2
    +96\alfor\lambda_4^2  \right.\nl
    \left.\quad\qquad
    -\frac{192}{5} g_5^2 M_5 -\frac{3}{5} g_X^2 M_X \right], \\
\frac{d \aone}{dt} & = & 
  \frac{1}{16 \pi^2}\left[ 96\aten\yten^2 + 4\afiv\yfiv^2 + 14\aone\yone^2 
    +96\alfor\lambda_4^2
    -\frac{96}{5} g_5^2 M_5 -\frac{19}{5} g_X^2 M_X \right], \\
\frac{d \alfor}{dt} & = & 
  \frac{1}{16 \pi^2}\left[ 96\aten\yten^2 + 2\aone\yone^2 + 288\alfor\lambda_4^2
    -\frac{192}{5} g_5^2 M_5 -\frac{3}{5} g_X^2 M_X \right], \\
\frac{d \alfiv}{dt} & = & 
  \frac{1}{16 \pi^2}\left[ 4\afiv\yfiv^2 + 4A_S y_S^2 + 288\alfiv\lambda_5^2
    -\frac{192}{5} g_5^2 M_5 -\frac{3}{5} g_X^2 M_X \right], \\
\frac{d A_S}{dt} & = & 
  \frac{1}{16 \pi^2}\left[ 96\aten\yten^2 + 2\afiv\yfiv^2 + 24A_S y_S^2
    +96\alfiv\lambda_5^2
    -\frac{144}{5} g_5^2 M_5 -\frac{1}{5} g_X^2 M_X \right].
\label{RGEtril}
\eea
Similarly to the MSSM, the bilinear terms decouple from the rest of FSU(5) RGEs, but we list them for completeness
\bea
\frac{d \mu_h}{dt} & = &
  \frac{\mu_h}{16 \pi^2} \left[ 48\yten^2 +2\yfiv^2 +\yone^2 +48\lambda_4^2 +48\lambda_5^2
    -\frac{48}{5} g_5^2 -\frac{2}{5} g_X^2 \right] \,, \\
\frac{d \mu_S}{dt} & = &
  \frac{\mu_S}{16 \pi^2} 20 y_S^2 \,, \\
\frac{d B_h}{dt} & = &
  \frac{1}{16 \pi^2} \left[ 96\aten\yten^2 +4\afiv\yfiv^2 +2\aone\yone^2 
    +96\alfor\lambda_4^2 +96\alfiv\lambda_5^2  \right.\nl
    \left.\quad\qquad
    -\frac{96}{5} g_5^2 M_5 -\frac{4}{5} g_X^2 M_X \right] \,, \\
\frac{d B_S}{dt} & = &
  \frac{1}{16 \pi^2} 40 A_S y_S^2 \,.
\label{RGEbil}
\eea

\section{The Renormalizations of SSB Parameters}
\label{sec:RGEs}

As a first step in analyzing the minimal CFSU(5) model, we illustrate some relevant features 
of the renormalization of the SSB parameters in the model. As was mentioned in the previous section,  FSU(5) naturally
 forces the neutrino Yukawa coupling to be equal to the top-quark Yukawa 
 coupling at the matching scale (see Eq.~\ref{matching1}). 
Such a large neutrino Yukawa coupling can have a significant effect on the sparticle spectrum of the MSSM, and thus
change the location of the regions of parameter space with an acceptable relic density~\cite{msugrarhn}. 
This is different from minimal SU(5) or the MSSM with the Type-1 seesaw, where RHN fields are  
added `by hand' and the neutrino
Yukawa coupling is free. In such scenarios, $h_\nu$ could be dialed to very small values,
thus making its effect on the sparticle spectrum unobservable. 
For our comparisons here, we therefore 
compare CFSU(5) to the CMSSM augmented by a Type-1 seesaw model with $h_\nu=h_t$ at the
unification scale (hereafter called the $\nu$CMSSM), rather than the more commonly discussed CMSSM
in which $h_\nu$ is assumed to be small or absent.

\begin{figure}
\epsfig{file=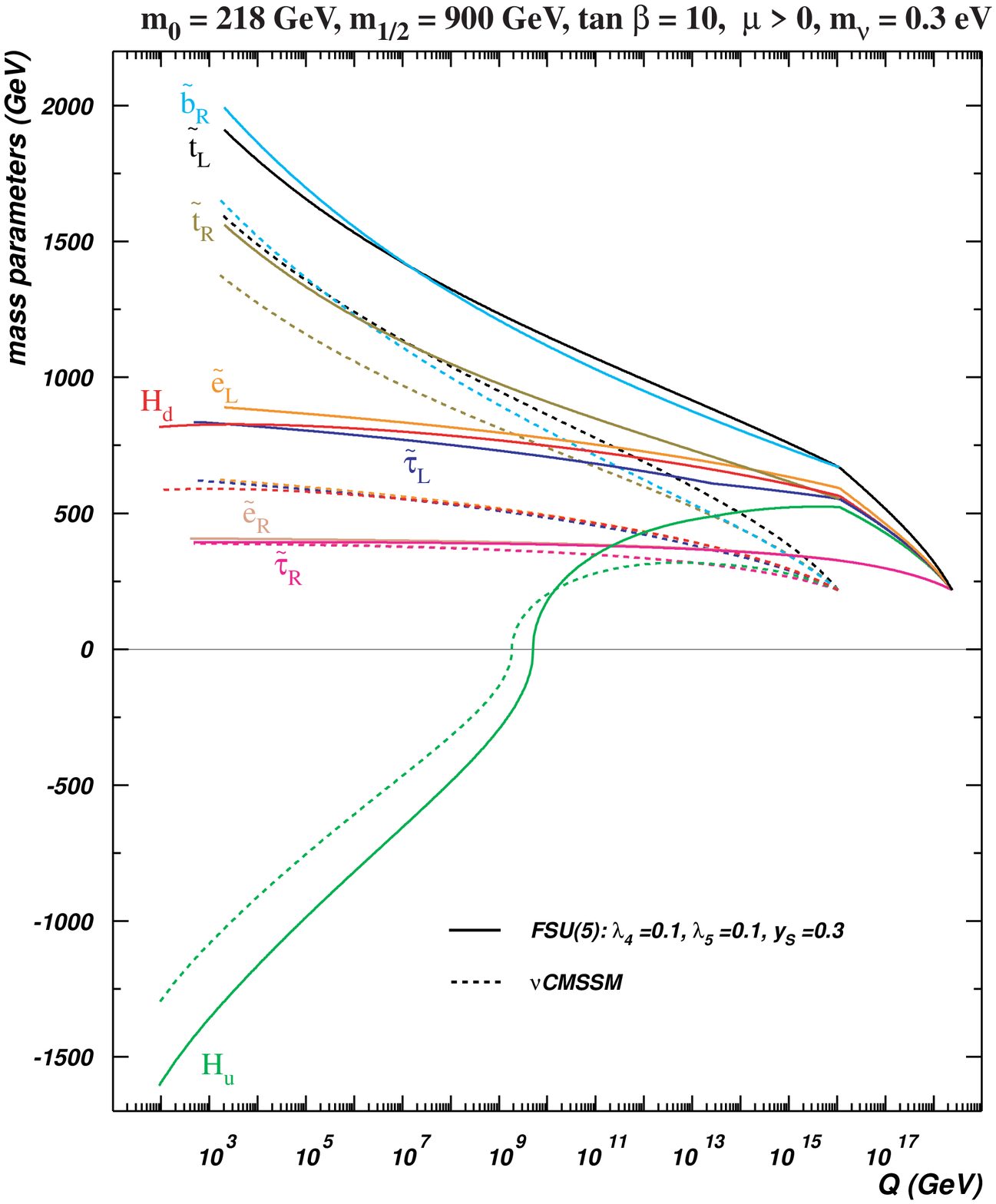,height=9.50cm}
\epsfig{file=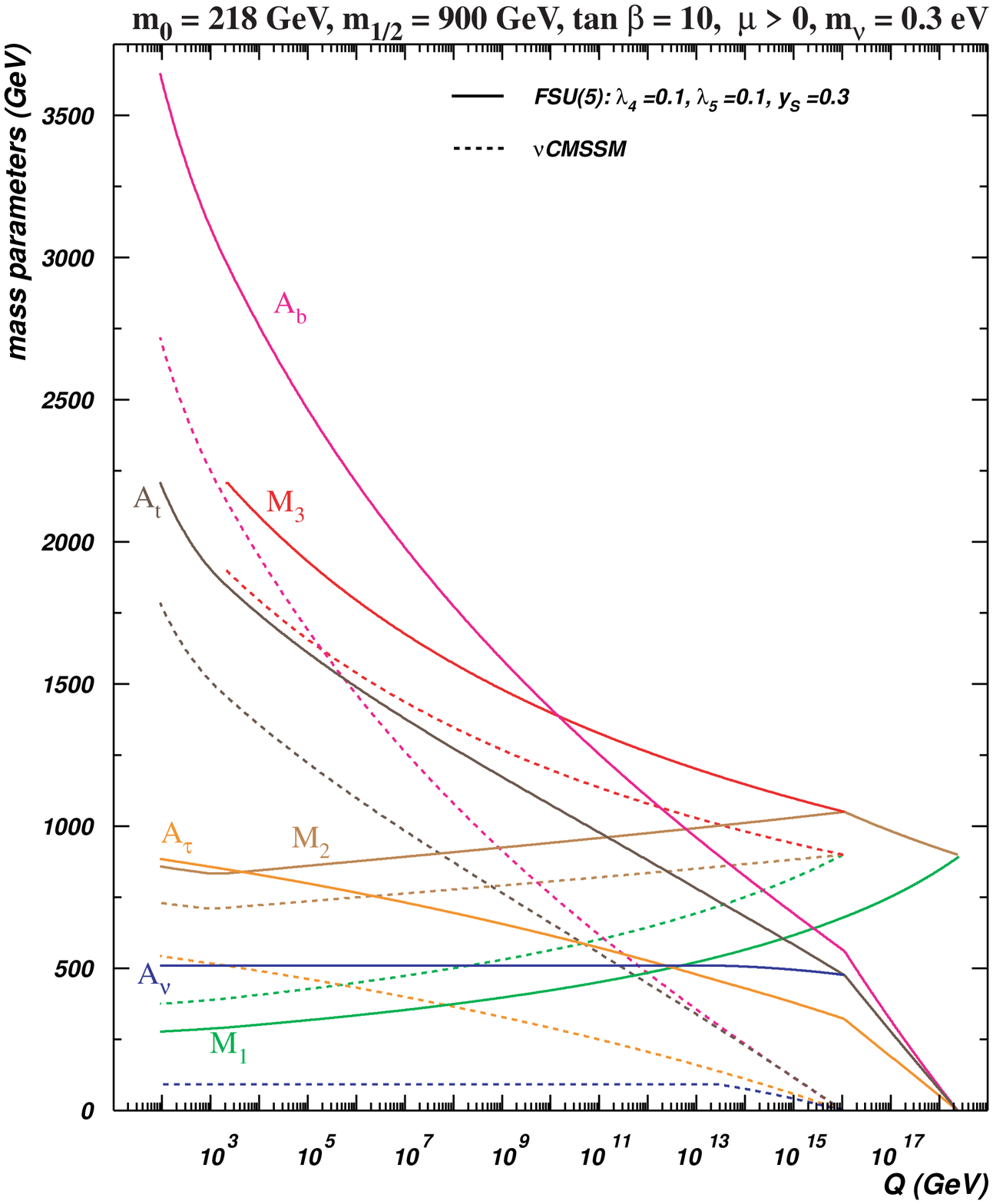,height=9.50cm}\\
\epsfig{file=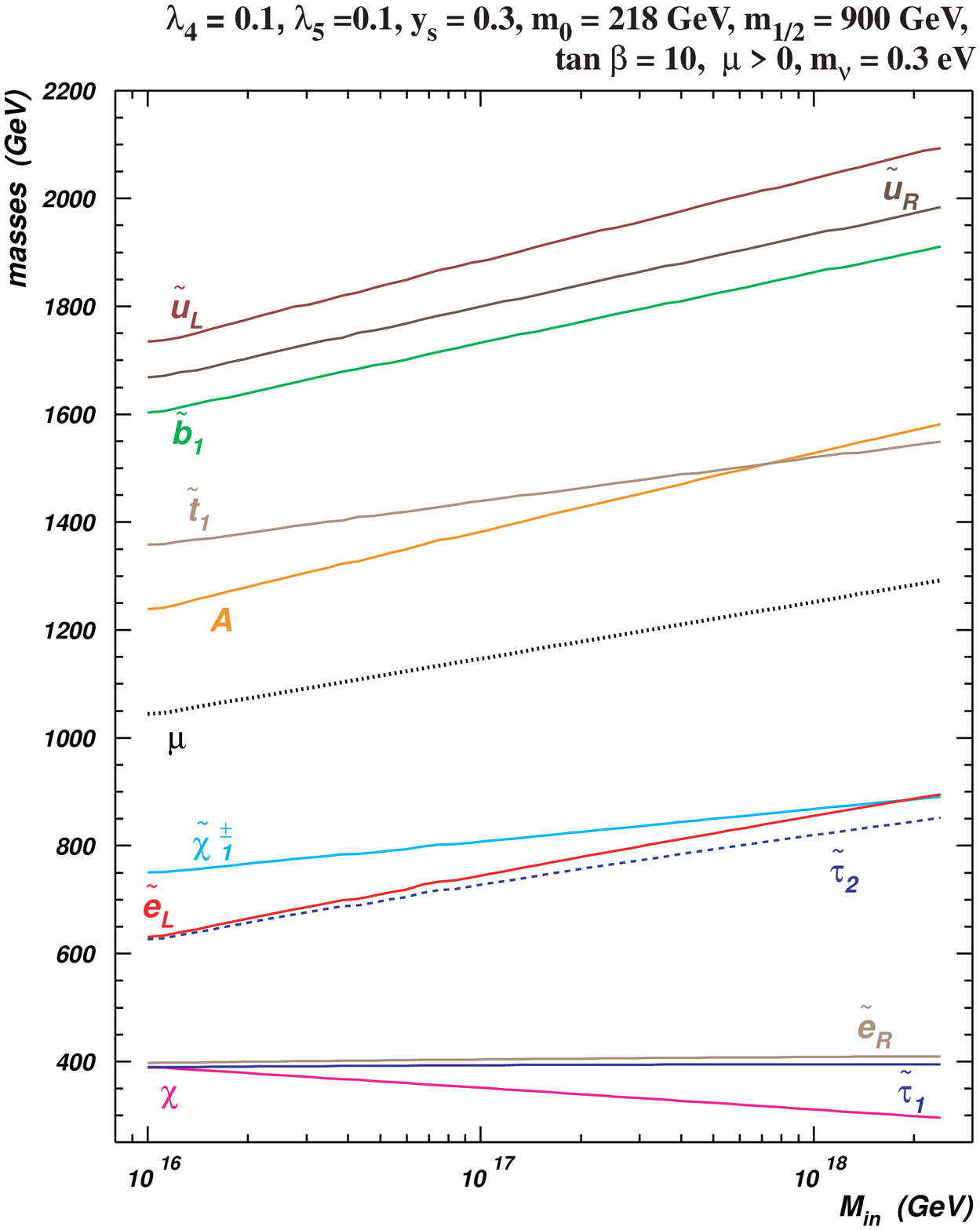,height=9.80cm}
\hskip .05in
\epsfig{file=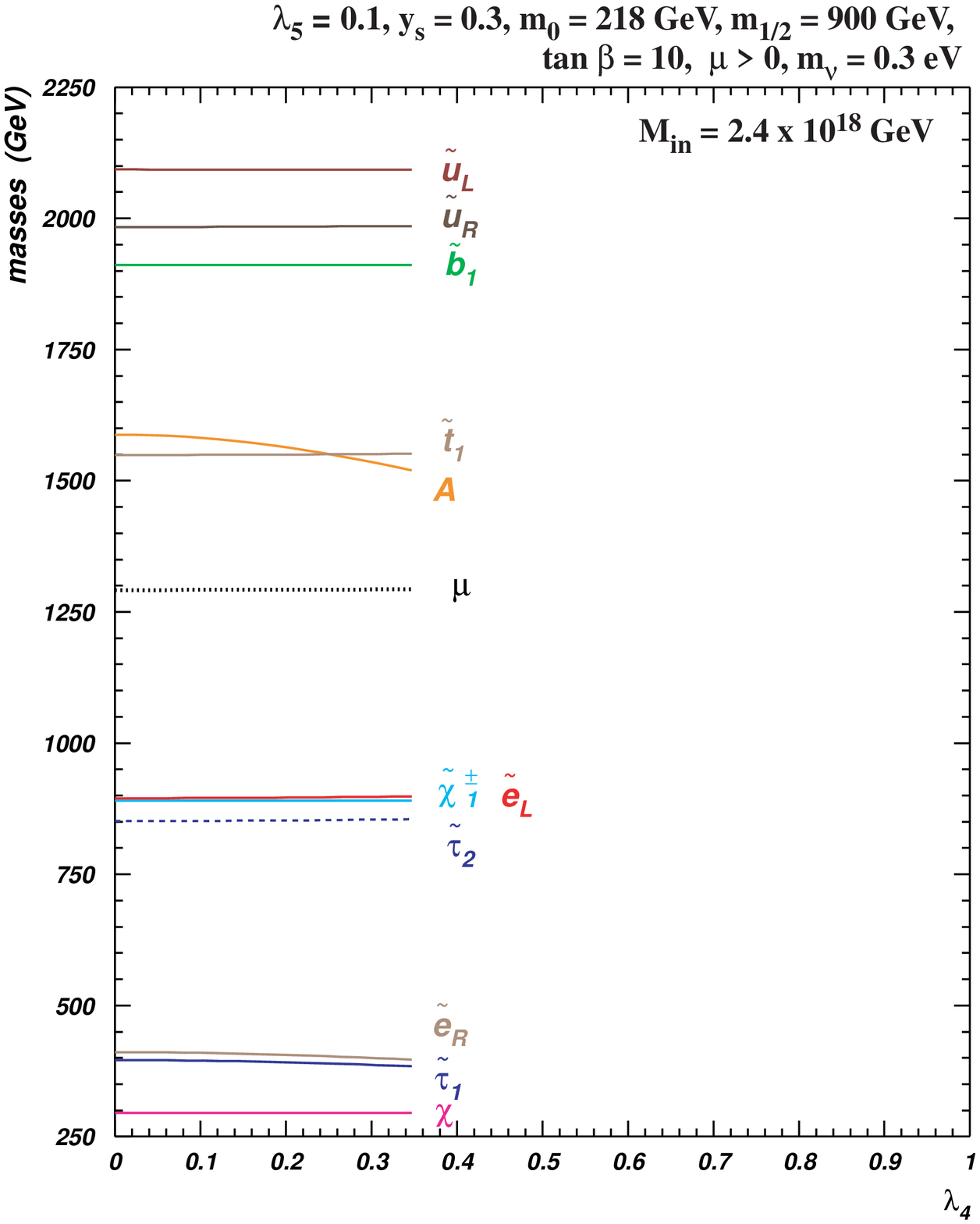,height=9.80cm}
\caption{\it RGE results for a stau coannihilation point $m_0 = 218$~GeV 
and $m_{1/2} = 900$~GeV, 
with $A_0 = 0$, $\lambda_4 = \lambda_5 = 0.1$, $y_S = 0.3$ at the input scale and $\tan \beta = 10$, $m_{\nu_3} = 0.3$~eV. 
Upper left: the evolution of SSB scalar mass parameters with choices $M_{in} =\mplr$ (solid lines) and $M_{in} =\mgut$ (dashed lines);
upper right: the evolution of gaugino masses and trilinear A-terms for $M_{in} =\mplr$ (solid lines) and $M_{in} =\mgut$ (dashed lines);
lower left: dependence of the physical sparticle/Higgs masses on $M_{in}$; 
lower right: dependence of the physical sparticle/Higgs masses on $\lambda_4(M_{in})$ assuming $M_{in}=\mplr$.
}
\label{fig:renn}
\end{figure}

In the upper left panel of Fig.~\ref{fig:renn} we compare the renormalizations of sfermion masses in the CFSU(5) model
(solid lines) with their renormalizations in the $\nu$CMSSM (dashed lines), for representative choices of the universal SSB parameters 
$m_0 = 218$~GeV, $m_{1/2} = 900$~GeV, $A_0 = 0$ at the input scale, and $\tan \beta = 10$.
In the $\nu$CMSSM,  these parameter choices specify a point close to the tip of the stau coannihilation strip~\cite{stauco} that reproduces the
cosmological density of cold dark matter. Furthermore, the choice of the neutrino Yukawa 
coupling and neutrino mass correspond to a right-handed neutrino mass $M_{N_3} = 2.8 \times 10^{13}$~GeV.
In CFSU(5), additional parameters are required to
fully specify the model, as listed in (\ref{CFSU5params}), and we make the choices 
$M_{in} =\mplr = 2.4 \times 10^{18}$~GeV, $\lambda_4 = \lambda_5 = 0.1$, $y_S = 0.3$. 
Note that our results are very insensitive to $y_S$, which we fix to 
0.3 throughout this paper. We see that the sfermion masses
are quite different already at the conventional GUT scale, though some converge again at lower scales,
{\it e.g.}, the ${\tilde e}_R, {\tilde \mu}_R$ and ${\tilde \tau}_1$ masses. On the other hand, some sfermion masses
remain quite different at low scales, {\it e.g.}, the ${\tilde e}_L, {\tilde \mu}_L$ and ${\tilde \tau}_2$ masses, while the
squark masses differ by $\sim 20$\% between the two models. In parallel, we note that the 
electroweak symmetry breaking, via the squared mass of the $H_u$ being driven negative,
arises qualitatively similarly in the two models.

The upper right panel of Fig.~\ref{fig:renn} shows the corresponding renormalizations
of the gaugino masses $M_a$ and the trilinear SSB parameters $A_i$. We see that both these
sets of quantities are quite different in the $\nu$CMSSM and CFSU(5). In particular, the ratios
$M_1/M_{2,3}$ are significantly smaller in CFSU(5), and the $A_i$ parameters are
significantly larger, thanks to the additional running between $M_{in}$ and $\mgut$.

\begin{figure}
\epsfig{file=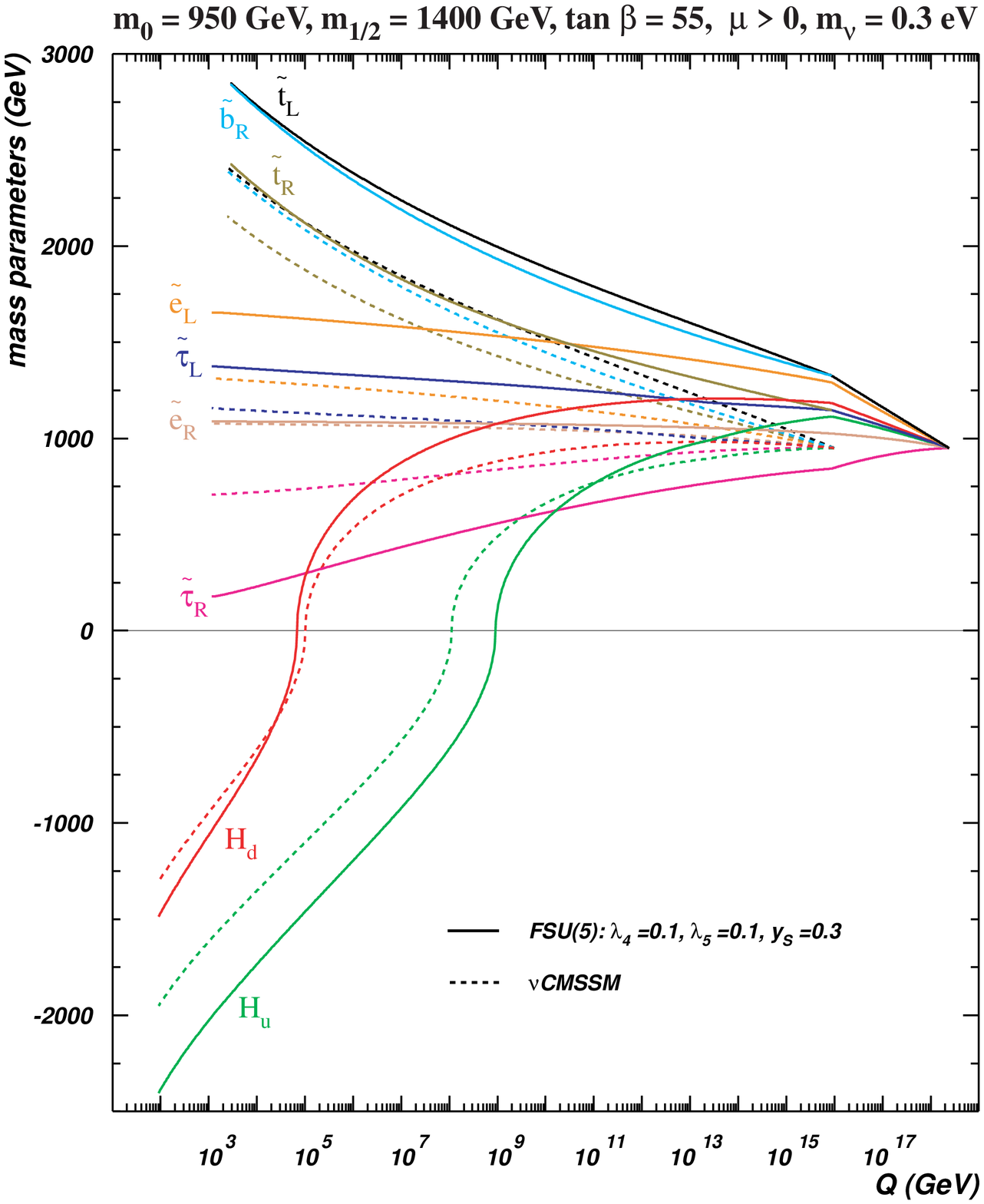,height=9.50cm}
\epsfig{file=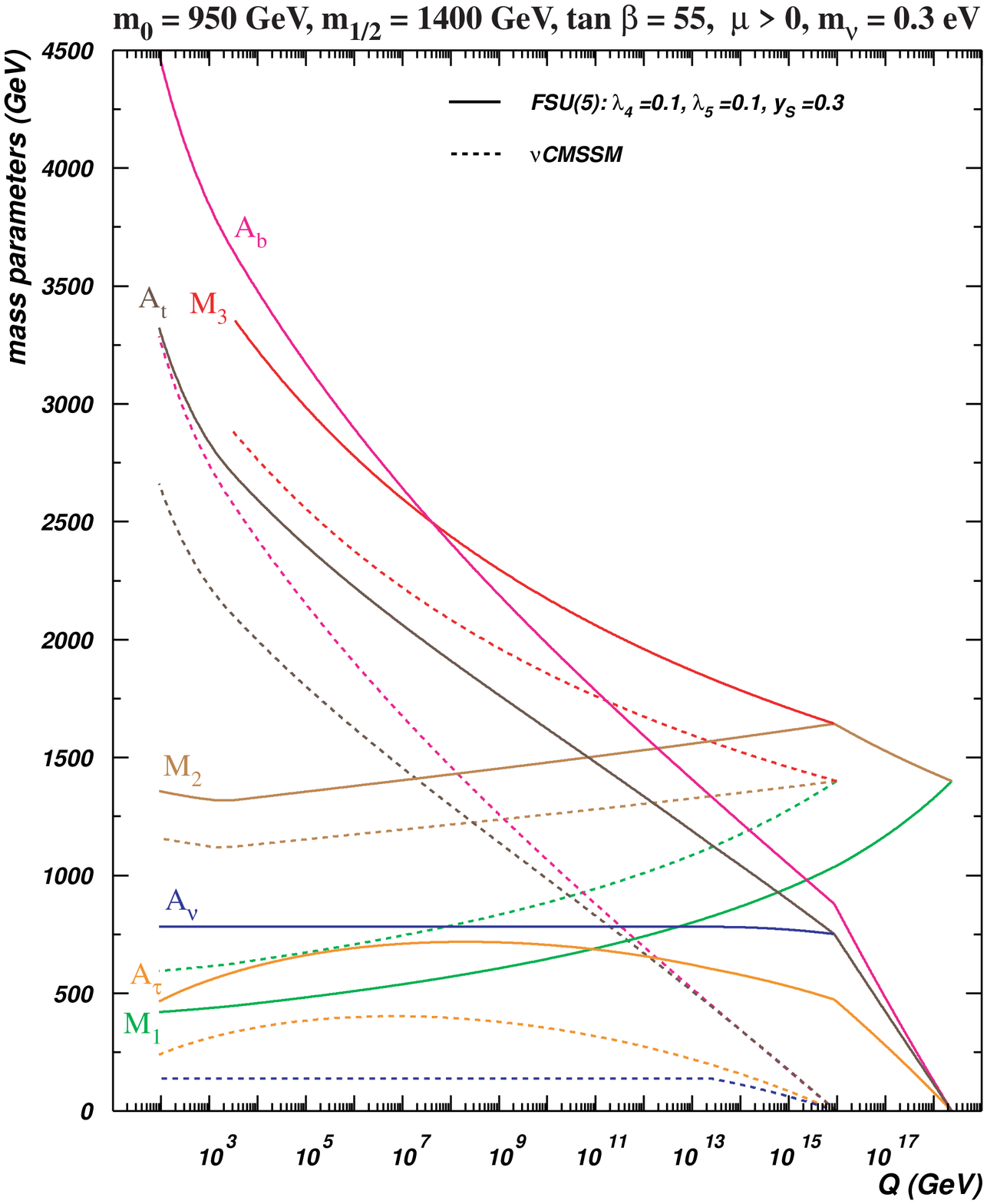,height=9.50cm}\\
\epsfig{file=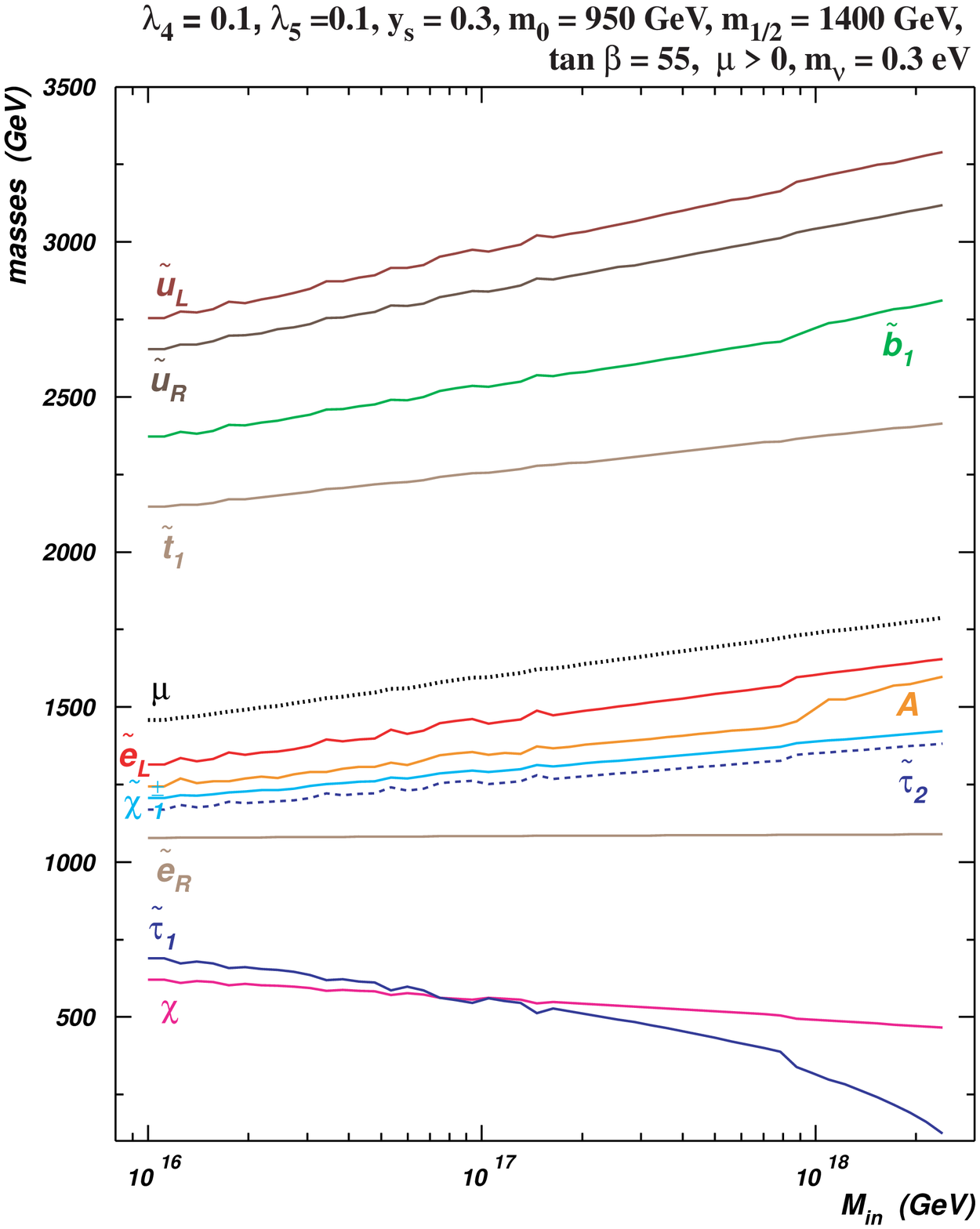,height=9.80cm}
\hskip .05in
\epsfig{file=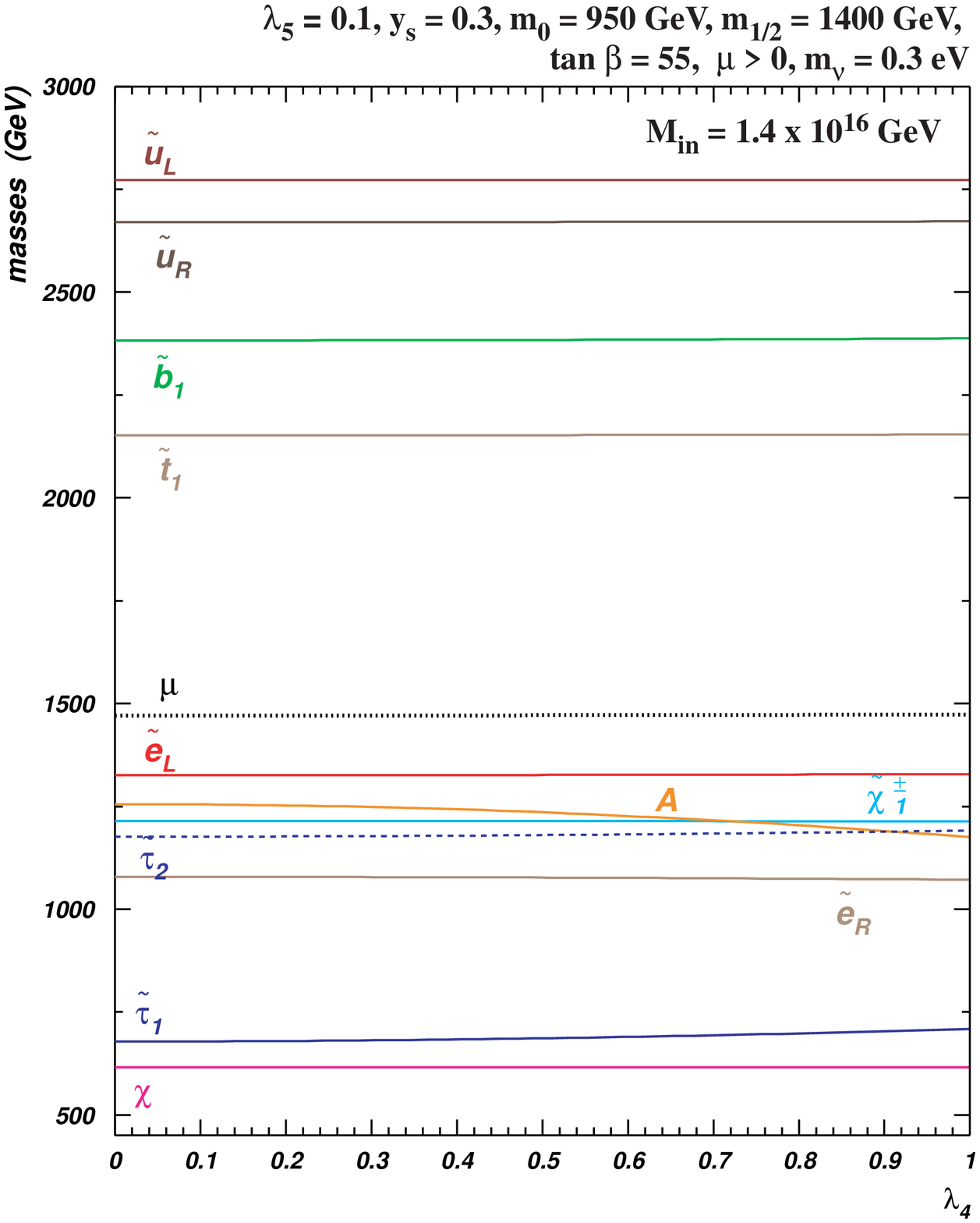,height=9.80cm}
\caption{\it
As for Fig.~\protect\ref{fig:renn}, but for $m_0 = 950$~GeV 
and $m_{1/2} = 1400$~GeV, $A_0 = 0$ at the input scale, and $\tan \beta = 55$.
}
\label{fig:renn55}
\end{figure}

Another aspect of these differences in the renormalizations is displayed in the lower left panel of
Fig.~\ref{fig:renn}, where $M_{in}$ is varied between $10^{16}$ and $2.4 \times 10^{18}$~GeV,
keeping fixed the other CFSU(5) model parameters: $\lambda_4 = \lambda_5 = 0.1$, $y_S = 0.3$. 
In line with our earlier remark, we see that the ${\tilde e}_R, {\tilde \mu}_R$ 
and ${\tilde \tau}_1$ masses are almost independent of $M_{in}$, 
As mentioned in Section~\ref{sec:fsu5}, right-handed neutrinos $N^c_i$ and singlinos $\tilde{S}_i$ have very large masses close to $\mgut$, and
therefore do not mix with MSSM higgsinos and gauginos. Thus FSU(5) neutralinos have the same compositions as those in the CMSSM~\cite{flippedDM}.
From Fig.~\ref{fig:renn} we see that the mass of the
lightest neutralino $\chi$ (which is almost pure bino) decreases with $M_{in}$, whilst the other sfermion masses and the mass
of the lighter chargino $\chi_1^\pm$ (which is dominantly wino) increase with $M_{in}$.

So far we considered one particular choice of the FSU(5) Yukawa couplings. 
The triple-Higgs Yukawa couplings have to be non-zero in order to achieve the 
desired doublet-triplet splitting, but otherwise are not
constrained. However, if we start from values of the Yukawa couplings at $\mgut$ that
are not very small, then they will quickly grow and might become
non-perturbative before the SSB unification scale is reached, as can be seen from Eqs.~(\ref{RGElam4}) and (\ref{RGElam5}). 
In the lower right panel of Fig.~\ref{fig:renn} we illustrate the sensitivity of the sparticle/Higgs spectrum to the choice of
$\lambda_4(\mgut)$. We terminate curves on the right when $\lambda_4(Q)$ reaches 5 before $M_{in}$;
at larger values of $\lambda_4(Q)$ the 2-loop contribution becomes comparable with the 1-loop part and our results could
become unreliable. 
We see that the sfermion masses are almost insensitive to $\lambda_4(\mgut)$, because 
the RGEs for the corresponding soft masses do not depend directly on the triple-Higgs Yukawa coupling as can be seen
from Eqs.~(\ref{RGEmass0}) - (\ref{RGEmass}). 
On the other hand, 
the mass of the CP-odd Higgs boson decreases as $\lambda_4(\mgut)$ increases, while $\mu$ remains constant. 
We can understand this behavior using approximate tree-level expressions for the relevant quantities.
For the moderate to large values of $\tan\beta$, that are favored by LEP Higgs boson mass constraints
and $|m_{H_u}|\gg M_Z$, we can write
\bea
\mu^2 &\simeq & -m^2_{H_u}(M_{weak})\, , \nnl
m^2_A & = & m^2_{H_u}(M_{weak})+m^2_{H_d}(M_{weak})+2\mu^2 \simeq m^2_{H_d}(M_{weak})-m^2_{H_u}(M_{weak}).
\label{eq:REWSB}
\eea
We see from Eqs.~(\ref{RGEmass0}) - (\ref{RGEmass}) and (\ref{matching2}) that larger $\lambda_4$ increases the
downward push in the $m_{h_1}^2$ RGEs that results in a smaller weak scale value of $m_{H_d}^2$. 
We also see that $m_{h_2}^2$ is not renormalized by $\lambda_4$ directly, so that
varying it has only a mild effect on the weak scale value of $m_{H_d}^2$. 
Consequently, the increase of $\lambda_4$ does not change the value of $\mu$, 
but the CP-odd Higgs boson becomes lighter. As
we show later, this has an important effect on the allowed regions. 
Making a similar analysis for $\lambda_5$, we find that increasing it makes both $\mu$ and $m_A$ larger,
with negligible effect on the sfermion masses. Therefore we do
not show the spectrum as function of $\lambda_5$, and fix $\lambda_5 =0.1$ for the rest of the paper. 

Another example of possible renormalization effects in CFSU(5) is shown in Fig.~\ref{fig:renn55},
which is similar to Fig.~\ref{fig:renn} apart from the choices $m_0 = 950$~GeV,
$m_{1/2} = 1400$~GeV and $A_0 = 0$ at the input scale, and $\tan \beta = 55$ corresponding to 
the rapid-annihilation funnel region~\cite{funnel,efgosi}. 
The corresponding RHN mass for these parameters is $M_{N_3} = 2.6 \times 10^{13}$~GeV. 
In this case, we see in the upper left panel that the ${\tilde e_R}$ and ${\tilde \mu_R}$
have similar masses in CFSU(5) and the $\nu$CMSSM, but not the ${\tilde \tau_1}$. This is because the
Yukawa renormalization effects are larger for $\tan \beta = 55$ than for $\tan \beta = 10$, 
so the $\yone^2$ term in Eq.~(\ref{RGEmassl}) dominates and pushes $m^2_l$ to smaller values, 
a feature visible already in the top left frame.  
We also see that the ${\tilde t_R}$ has a similar mass
in both models, which is due to a compensation between renormalization effects above $\mgut$
in the CFSU(5) case and different renormalizations at $Q < \mgut$ in the two models. We see in the
upper right panel of Fig.~\ref{fig:renn55} that the renormalizations of all gaugino masses and trilinear
SSB parameters are different in the two models. In the lower left panel, we note in particular the level
crossing between the mass of the lightest neutralino $\chi$ and the lighter stau ${\tilde \tau_1}$, and
also that the ratio $m_\chi/m_A$ decreases monotonically as $M_{in}$ increases. Finally,
in the lower right panel of Fig.~\ref{fig:renn55} we see that only $m_A$ 
is very sensitive to $\lambda_4$, and the same is true for $\lambda_5$ (not shown).

\begin{figure}[ht]
\begin{center}
\epsfig{file=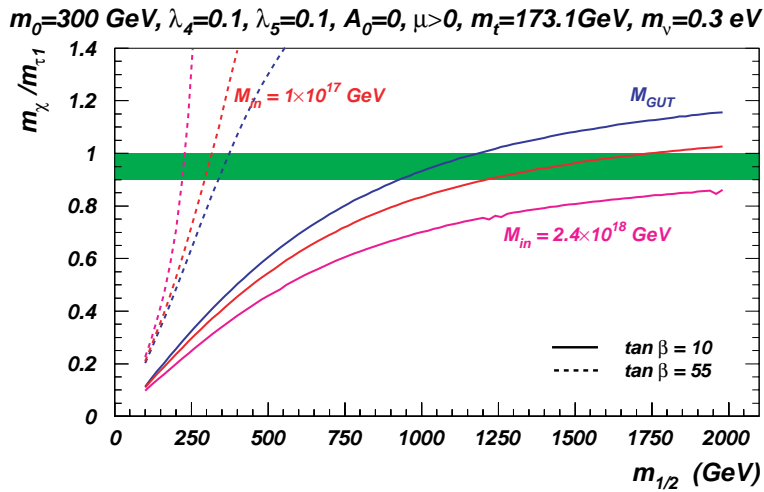,width=10cm}
\end{center}
\vskip -.2in
\caption{\it
The ratio of the masses of the lightest neutralino $\chi$ and the lighter stau ${\tilde \tau}_1$ as a 
function of $m_{1/2}$ for the three choices $M_{in} = \mgut, 10^{17}$~GeV and $\mplr$,
assuming $m_0 = 300$~GeV, $A_0 = 0$ and $\tan \beta = 10$ (solid) or 55 (dashed), 
and $\lambda_4 = 0.1, \lambda_5 = 0.1$, $y_S = 0.3$.  The shaded green horizontal band highlights the regime 
in which stau coannihilation is important.}
\label{fig:chi2stau}
\end{figure}

We focus in Fig.~\ref{fig:chi2stau} on the ratio of the $\chi$ and ${\tilde \tau}_1$ masses,
as a function of $m_{1/2}$ for representative choices of $M_{in}$ and $\tan \beta$, and 
fixing the other CFSU(5) parameters
to be $m_0=300$~GeV, $\lambda_4 = \lambda_5 = 0.1$, $A_0 = 0$ and $m_{\nu_3} = 0.3$~eV.
The blue curves correspond to the case $M_{in} = \mgut$, where CFSU(5) reduces to the
$\nu$CMSSM. 
The coannihilation processes become important when the mass gap between the LSP and the NLSP is 
$\lesssim 15\%$~\cite{Griest:1990kh}; this regime is indicated by the horizontal green band.
Within this band, compatibility with WMAP is achieved in a narrower range of the mass ratio that is model-dependent. 
We see that in the $\nu$CMSSM the $\chi$ and ${\tilde \tau}_1$ masses approach close enough for
coannihilation to become important, bringing the relic density into the WMAP range, for
$m_{1/2} \sim 1000 (350)$~GeV when $\tan \beta = 10 (55)$ and $m_0 = 300$~GeV.
Since the presence of the RHN and $h_\nu$ have only a modest effect on the stau
co-annihilation region~\cite{msugrarhn}, this result is also found in the CMSSM.
 On the other hand, when
$M_{in}$ increases in CFSU(5), we see that for $\tan \beta = 10$ the coannihilation region recedes to
larger values of $m_{1/2}$, as could be expected on the basis of Fig.~\ref{fig:renn} by
comparing the ${\tilde \tau}_1$ and $M_1$ in the upper panels, or by looking directly at the lower left panel.
However, when $M_{in}$ increases in CFSU(5) for $\tan \beta = 55$, the position of the
coannihilation region instead moves to {\it lower} $m_{1/2}$ due to the RGE running effect of $m^2_l$ described in the previous
paragraph. These results indicate that the
regions of the $(m_{1/2}, m_0)$ planes where coannihilation makes the relic $\chi$ density
compatible with WMAP are likely to be quite different in CFSU(5) from what they would be in the CMSSM.

\begin{figure}[ht]
\begin{center}
\epsfig{file=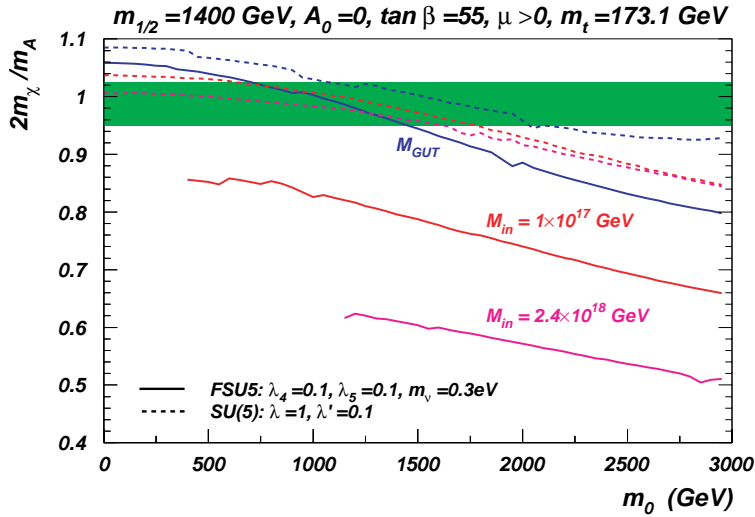,width=10cm}
\end{center}
\vskip -.2in
\caption{\it
The ratio of the masses of the lightest neutralino $\chi$ and the CP-odd Higgs boson $A$ as a function of $m_0$
for the three choices $M_{in} = \mgut,10^{17}$~GeV and $\mplr$, assuming
$m_{1/2} = 1400$~GeV, $A_0 = 0$ and $\tan \beta = 55$. We compare the cases
of minimal conventional SU(5) (dashed lines) assuming $\lambda = 1, \lambda^\prime = 0.1$ 
and CFSU(5) (solid lines) assuming $\lambda_4 = 0.1, \lambda_5 = 0.1$, $y_S = 0.3$. The shaded green 
horizontal band highlights the region in which rapid annihilation 
through the heavy Higgs funnel is important. Curves are terminated when $\tilde{\tau_1}$ becomes the LSP.}
\label{fig:chi2ma}
\end{figure}

Another prominent feature in $(m_{1/2}, m_0)$ planes in the CMSSM and $\nu$CMSSM is the appearance of a funnel
at large $\tan \beta$, where rapid $\chi \chi$ annihilation through direct-channel heavy Higgs ($A, H$)
resonances brings the relic density into compliance with WMAP~\cite{funnel,efgosi}. 
This funnel is very sensitive to relative masses of neutralino and heavy Higgs bosons, 
appearing when $m_\chi \sim m_{A,H}/2$. As can be seen in the lower panels of
Fig.~\ref{fig:renn55}, the ratio $m_\chi/m_{H,A}$ is very sensitive to $M_{in}$ (left) 
and to $\lambda_4$ (right). 
Fig.~\ref{fig:chi2ma} displays this ratio as a function of $m_0$ in minimal
conventional SU(5) (dashed lines) 
and CFSU(5) (solid lines) assuming $m_{1/2} = 1400$~GeV, $A_0 = 0$, $m_{\nu_3} = 0.3$~eV
and $\tan \beta = 55$ and with the choices $M_{in} = \mgut, 10^{17}$~GeV and $\mplr$.
In the CFSU(5) case, it is assumed that $\lambda_4 = 0.1$ and $\lambda_5 = 0.1$, whereas in the minimal
conventional SU(5) case it is assumed that $\lambda = 1$ and $\lambda^\prime = 0.1$ in the
notation of~\cite{EMO}. The horizontal green band in Fig.~\ref{fig:chi2ma} indicates where rapid
annihilation via the heavy Higgs funnel takes place. 
We see that at $M_{in} = \mgut$ the funnel is located at relatively similar but not identical values of $m_0$. 
This is due to the effect of the large
neutrino Yukawa coupling in $\nu$CMSSM (solid blue line) that increases $m_A$ consequently shifting the heavy Higgs funnel location to
lower $m_0$, as compared to CMSSM (dashed blue lines)~\cite{msugrarhn}.
The rapid-annihilation funnel feature looks very different for $M_{in} > \mgut$. Comparing to minimal conventional SU(5), where the funnel was
present for all values of $M_{in}$, we see that in the CFSU(5) funnel disappears very rapidly. 
As was shown earlier (see the discussion for Fig.~\ref{fig:renn55}) the FSU(5) RGEs drive the neutralino and heavy Higgs boson masses in opposite
directions, and the resonance regime disappears very rapidly with growing $M_{in}$. In contrast, in minimal SU(5) 
both $m_\chi$ and $m_A$ were
growing with $M_{in}$ and thus the resonant condition can always be achieved~\cite{EMO}.

\section{Representative ($m_{1/2},m_0$) Planes}
\label{sec:planes}

\begin{figure}
\epsfig{file=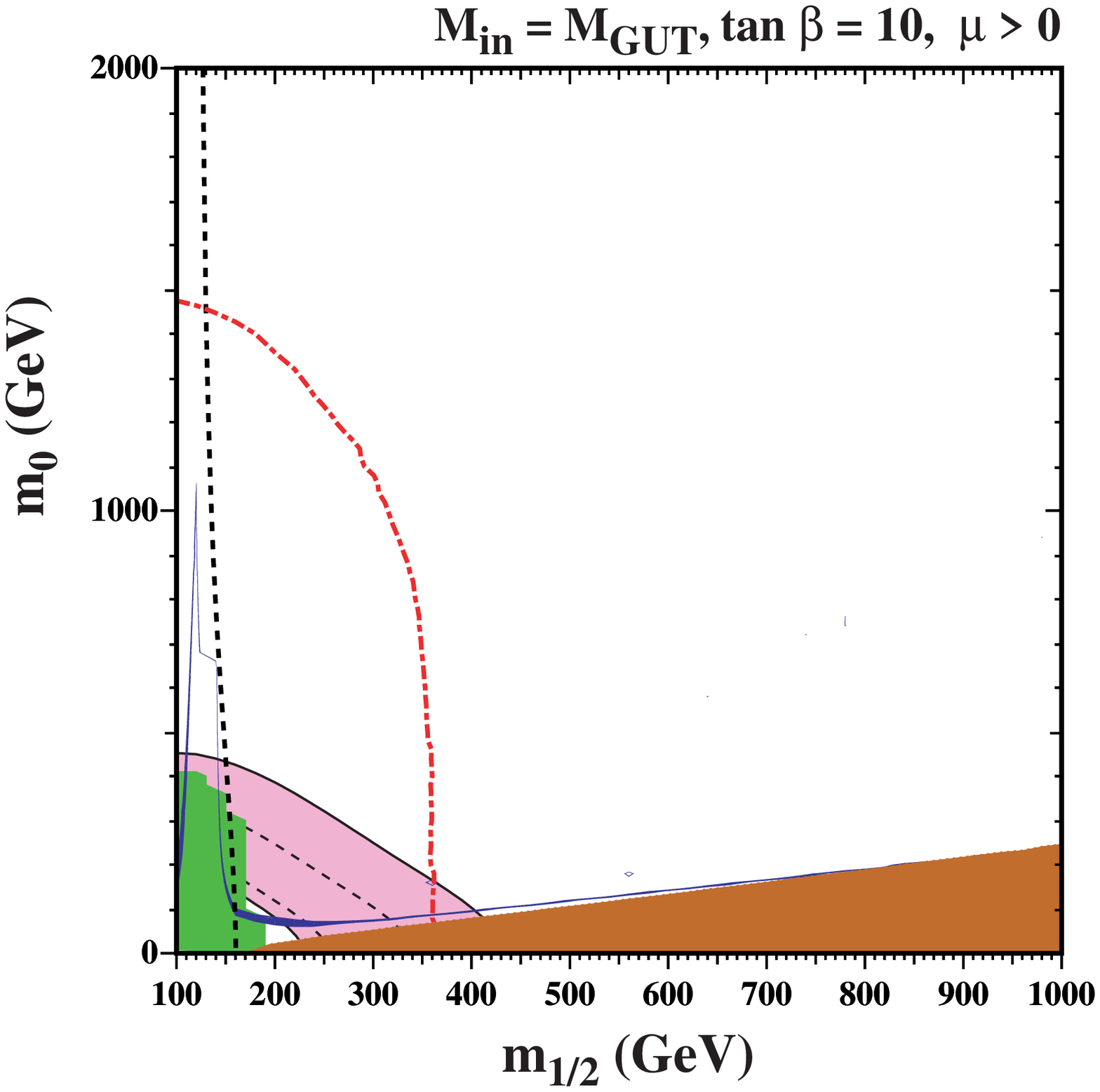,height=8.0cm}
\epsfig{file=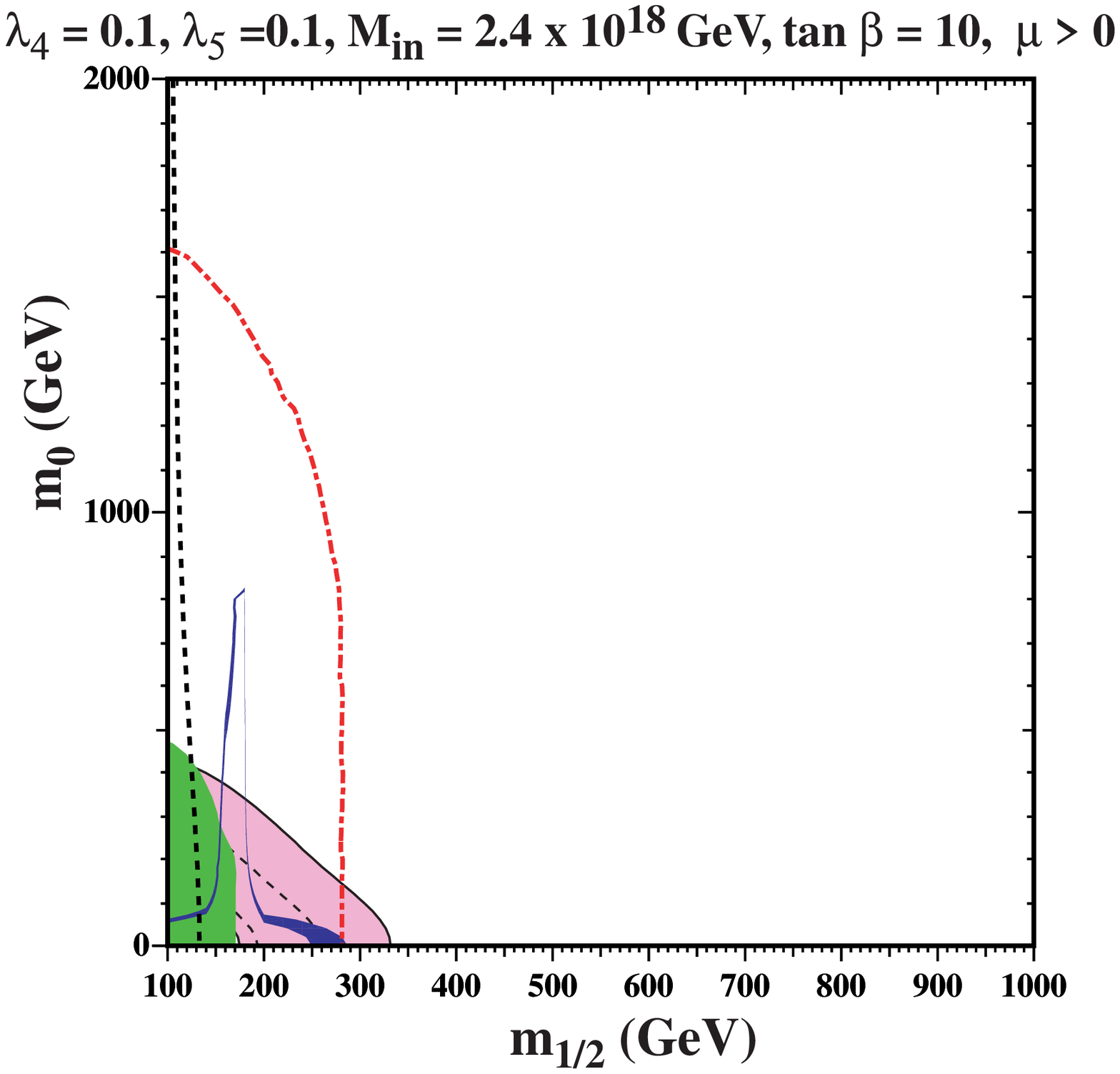,height=8.0cm}
\caption{\it
The $(m_{1/2}, m_0)$ planes for the $\nu$CMSSM (left) and for the minimal CFSU(5) model
with $A_0 = 0, \tan \beta = 10, \lambda_4 = 0.1$
and $\lambda_5 = 0.1$ for $M_{in} = 2.4 \times 10^{18}$~GeV (right). 
In the blue regions, $\ohsq$ is within the WMAP range. The
pink regions between the black dashed (solid) lines are allowed by $g_\mu-2$ at the 1-$\sigma$ (2-$\sigma$)
levels. The brown and green regions are excluded because the LSP is the ${\tilde \tau_1}$ and by
measurements of $b \to s \gamma$, respectively.
Areas to the left of the black dashed and red dash-dotted lines are ruled out by LEP searches 
for charginos and the lightest MSSM Higgs boson $h$, respectively. More details can be found in the text.}
\label{fig:tb10}
\end{figure}

Following these illustrative studies of renormalization effects in CFSU(5), we now examine a few
representative ($m_{1/2},m_0$) planes for $\tan \beta = 10$ and 55. In Fig.~\ref{fig:tb10} we
compare the $\nu$CMSSM case (left) with the CFSU(5) model for $\lambda_4 = 0.1, \lambda_5 =0.1$
and $M_{in} = 2.4 \times 10^{18}$~GeV (right). 
Note that, for $\tan \beta = 10$, the $(m_{1/2},m_0)$ plane in the $\nu$CMSSM is very similar 
to the CMSSM with no neutrino masses for values of $m_0$ below the focus-point region~\cite{fp}
(which is not visible in the left panel of Fig.~\ref{fig:tb10} for the current choice of parameters). 
In the brown region in the left panel, the ${\tilde \tau_1}$
would be the LSP, which is not allowed by astrophysics: there is no corresponding region in the
right panel. The green regions are disallowed by experimental measurements~\cite{hfag} of $b \to s \gamma$
decay~\footnote{The shaded region is excluded at the 95\% CL following the procedure of Ref.~\cite{bsgprocedure} and using the code by Gambino and
Ganis~\cite{gam}.}, 
and LEP limits on the masses of the lighter chargino~\cite{LEPsusy} and the lightest Higgs boson $h$~\cite{LEPHiggs} forbid
areas to the left of the black dashed and red dot-dashed lines, respectively. The pink regions are
favoured by $g_\mu - 2$ at the 1-$\sigma$ (2-$\sigma$) levels~\cite{newBNL}, as indicated by the dashed (solid)
lines. The dark blue strips are where the relic $\chi$ density falls within the range allowed by WMAP
and other experiments~\cite{WMAP}. In the left panel, for the $\nu$CMSSM, we see a well-developed coannihilation strip, 
which is curtailed in the right panel, for CFSU(5), and only marginally compatible with the LEP Higgs
constraint~\footnote{We recall that in the corresponding minimal conventional SU(5) case~\cite{EMO} the
remnant of the coannihilation strip lies entirely inside the region forbidden by the LEP constraint
on $m_h$.}. In both panels, we see a vertical funnel due to rapid annihilation through the direct-channel
$h$ pole~\cite{hfunnel}. This region has moved to larger $m_{1/2}$ because of the reduction in $m_\chi$
due to the extra CFSU(5) renormalization of $M_1$ 
between $M_{in}$ and $\mgut$, whose effects are visible in the upper right and lower left panels of
Fig.~\ref{fig:renn} (in that case for a different value of $m_{1/2}$). However, despite this extra
renormalization, this rapid-annihilation funnel is still in the region of the $(m_{1/2}, m_0)$ plane
forbidden by the LEP Higgs constraint.

\begin{figure}
\epsfig{file=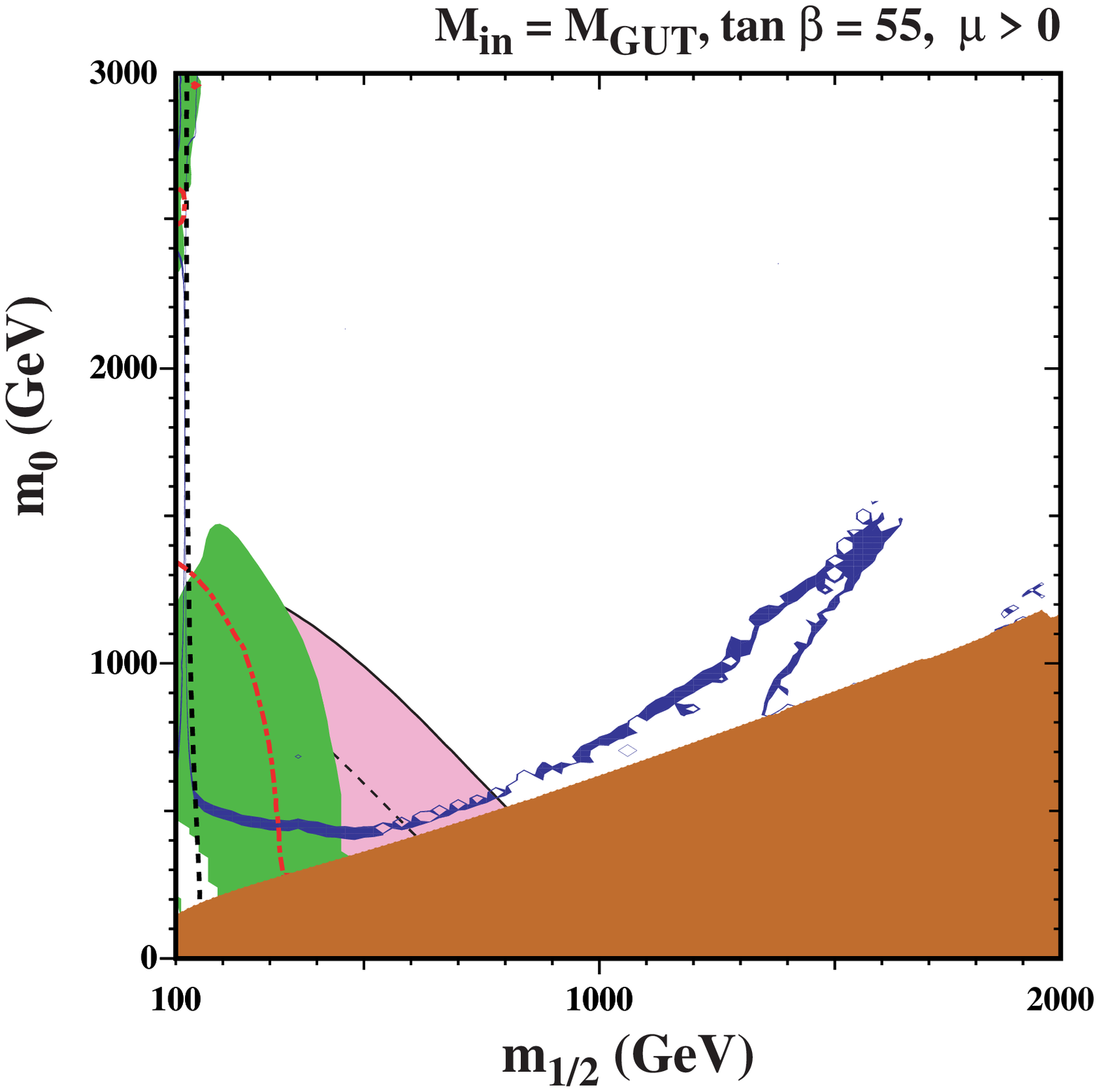,height=8.0cm}
\epsfig{file=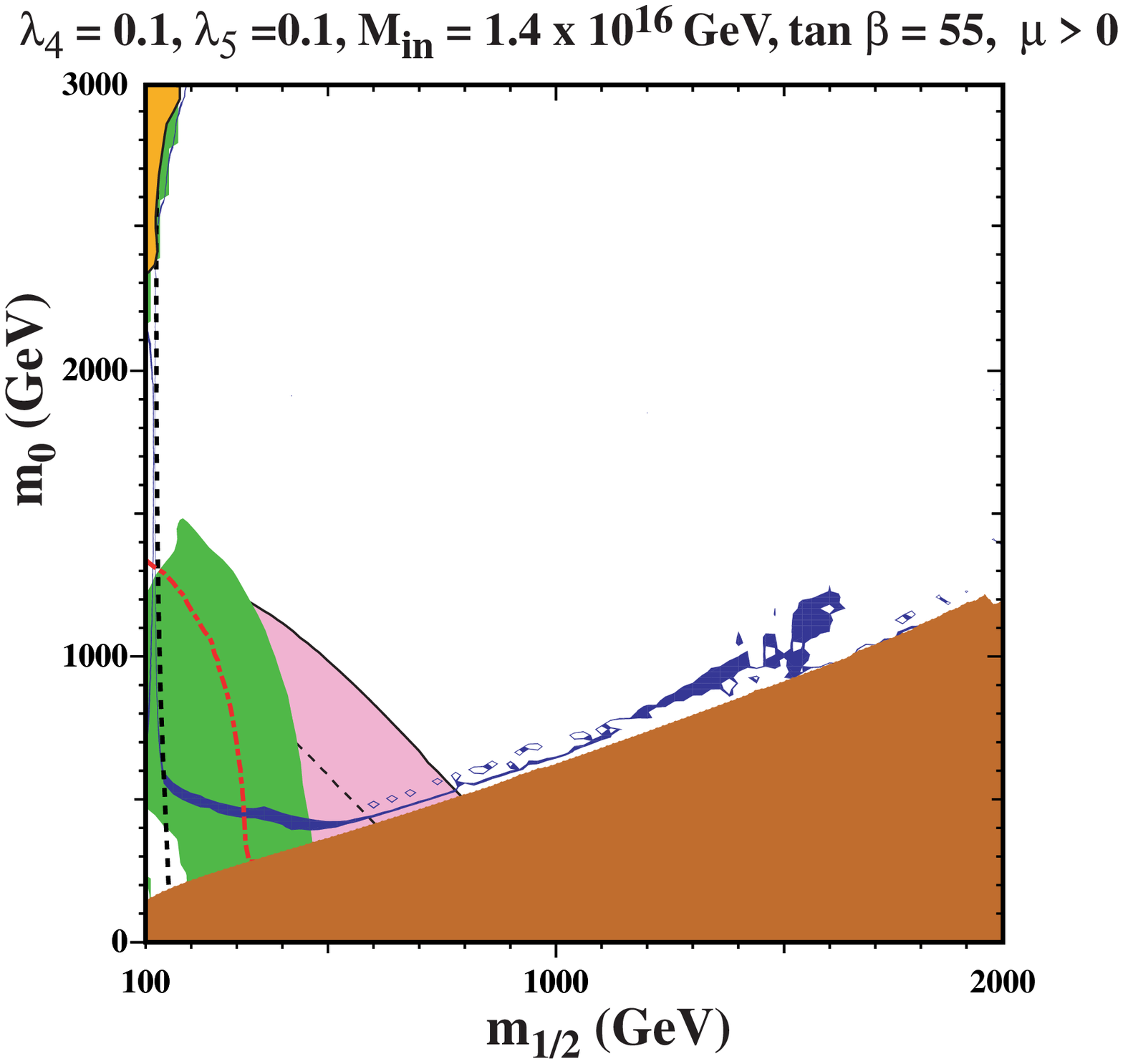,height=8.0cm}\\
\epsfig{file=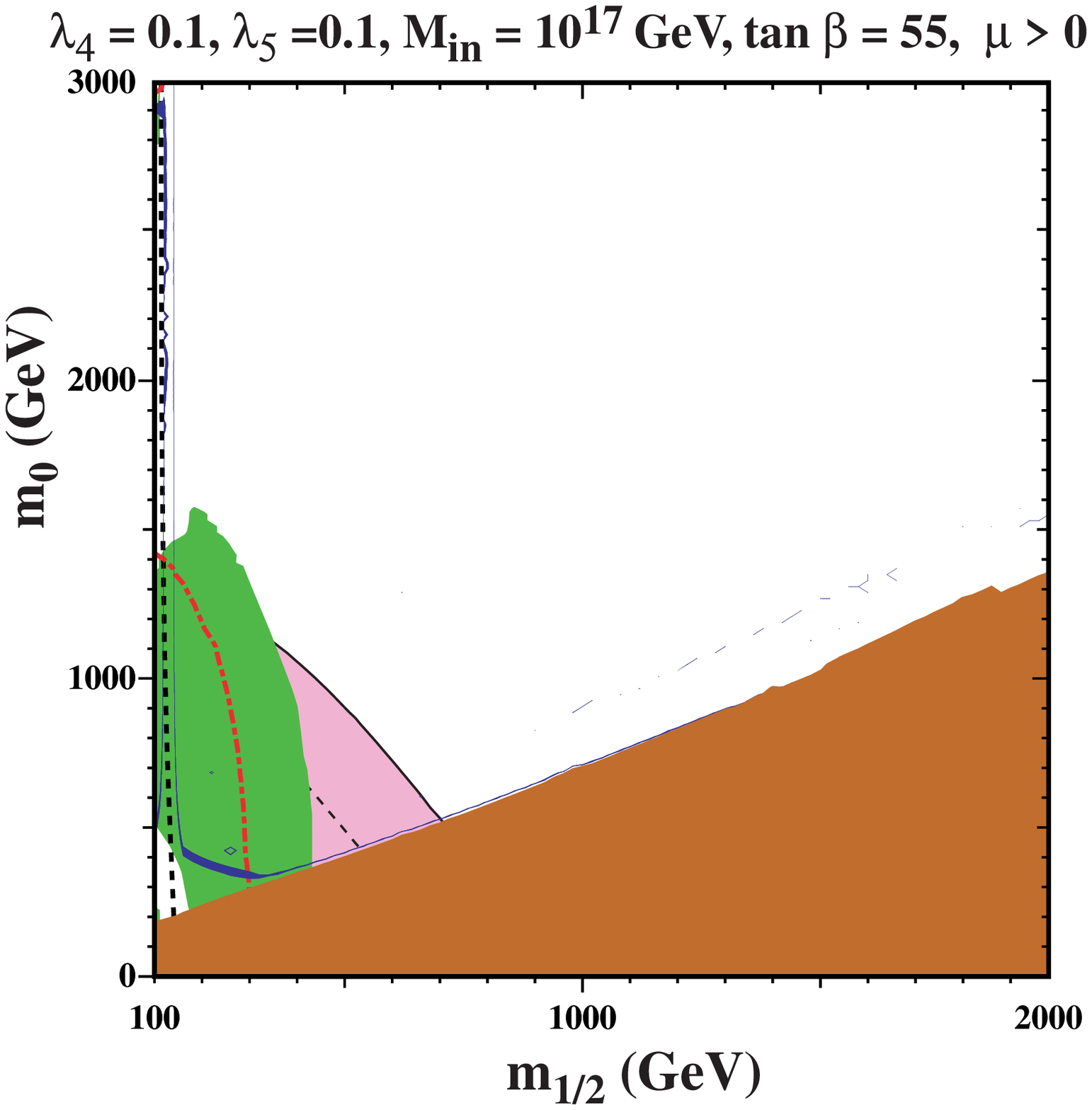,height=8.0cm}
\epsfig{file=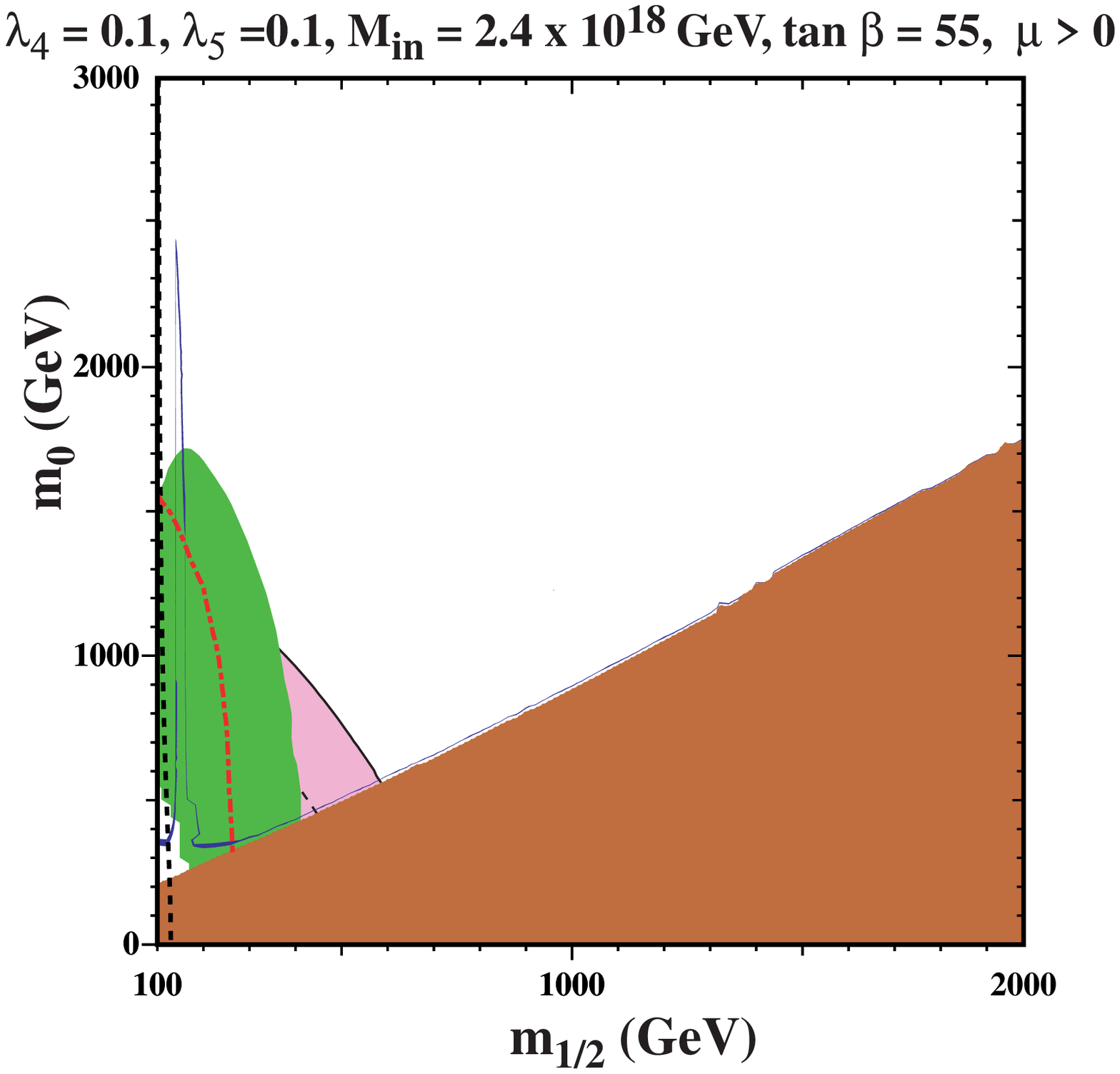,height=8.0cm}
\caption{\it
As for Fig.~\protect\ref{fig:tb10}, but for $\tan \beta = 55, \lambda_4 = 0.1$
and $\lambda_5 = 0.1$ with different choices of $M_{in}$:
$\mgut$ (upper left), $1.4 \times 10^{16}$~GeV (upper right ), $10^{17}$~GeV (lower left), 
and $2.4 \times 10^{18}$~GeV (lower right). 
}
\label{fig:tb55}
\end{figure}

\begin{figure}
\epsfig{file=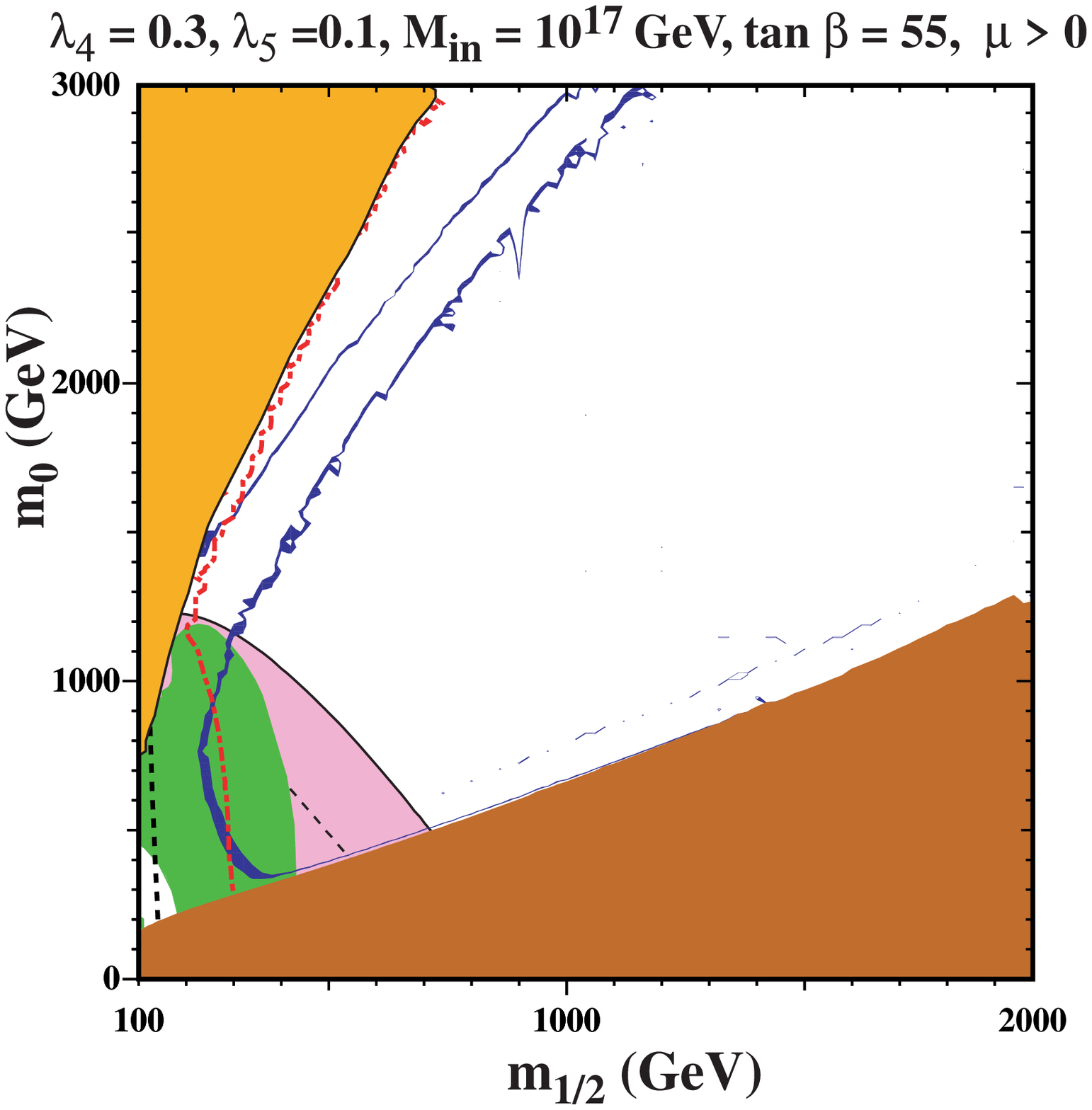,height=8.0cm}
\epsfig{file=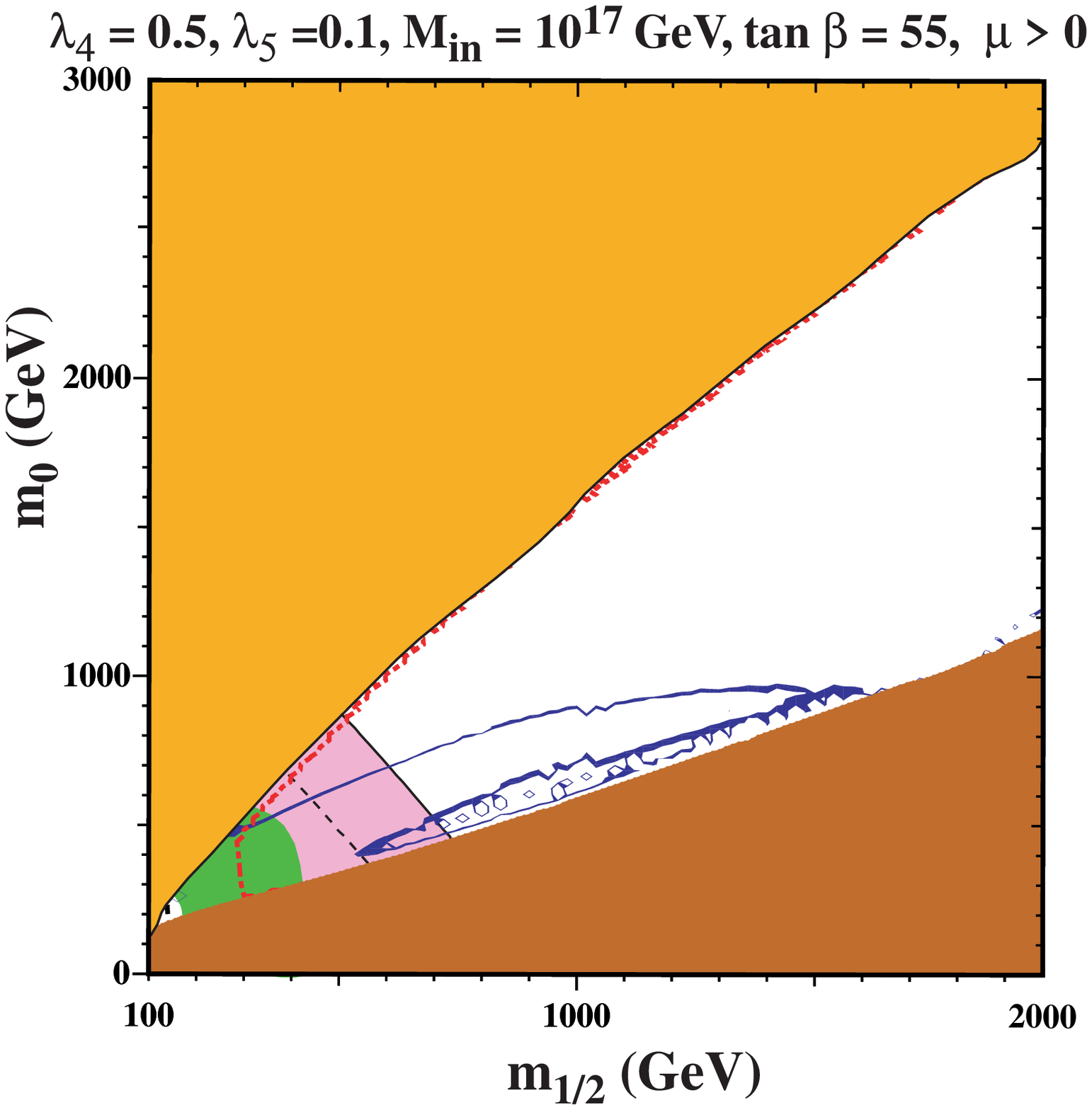,height=8.0cm}\\
\epsfig{file=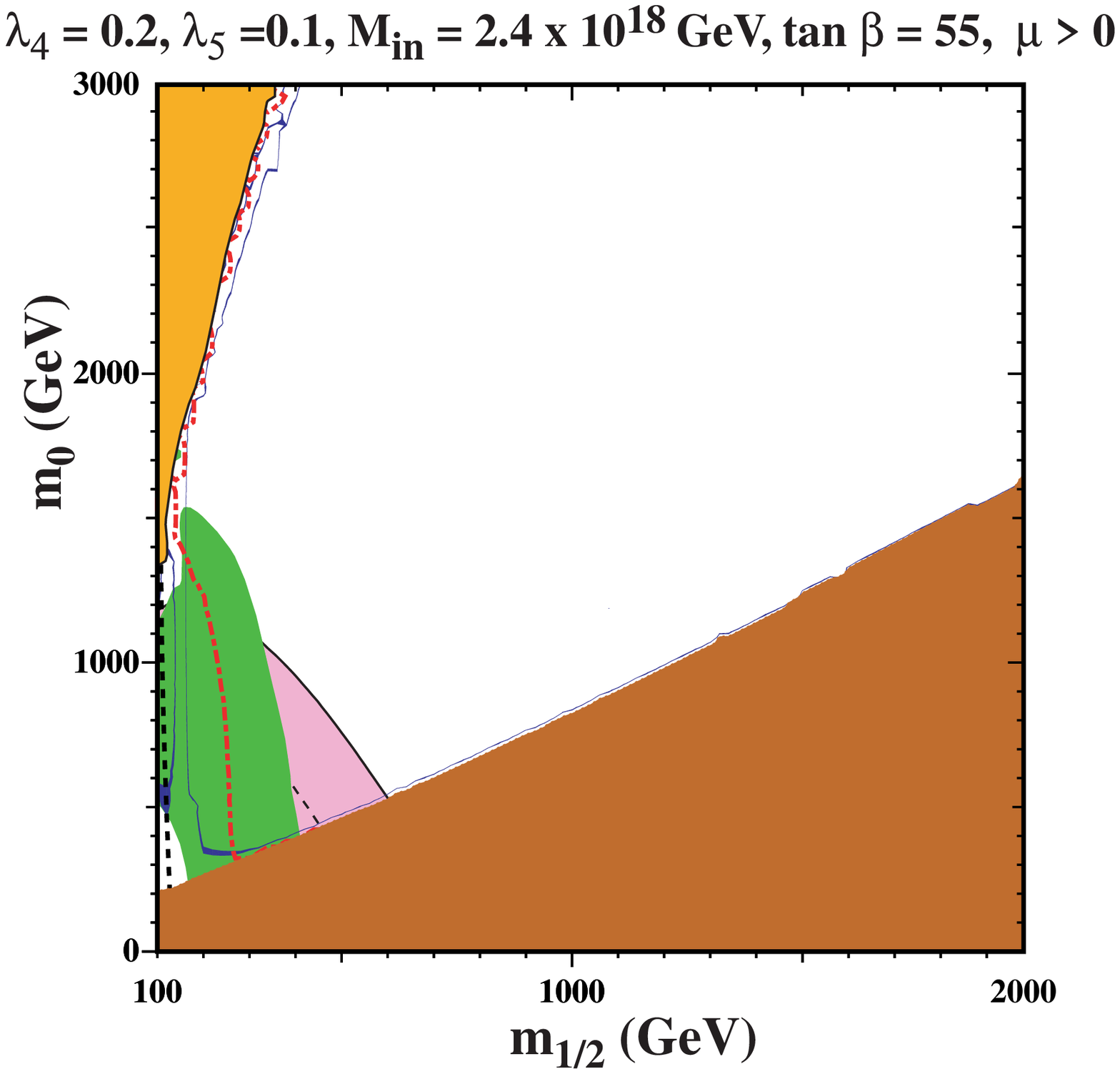,height=8.0cm}
\epsfig{file=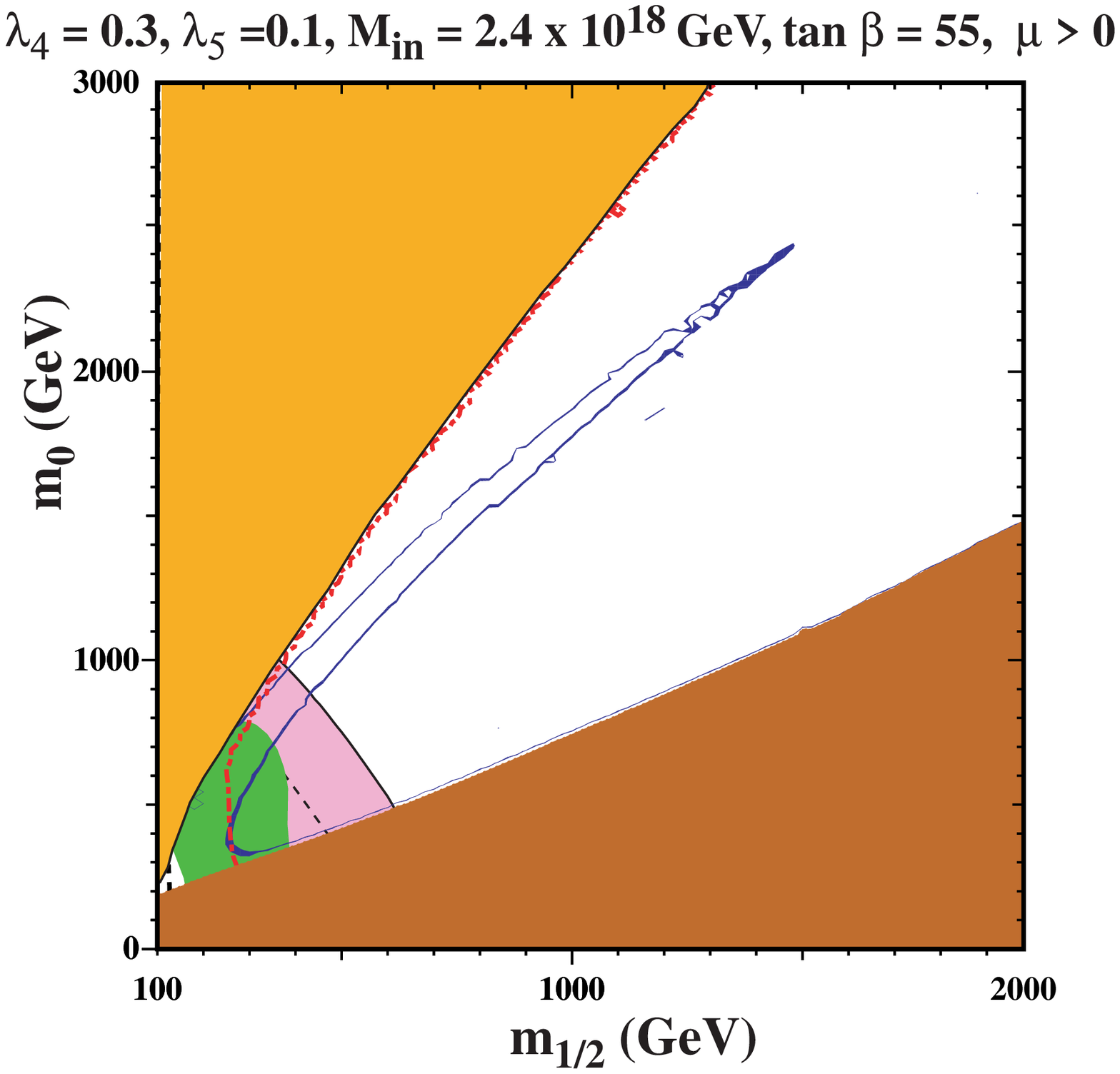,height=8.0cm}
\caption{\it
As for Fig.~\protect\ref{fig:tb10}, for $\tan \beta = 55$ and $\lambda_5 =0.1$, 
with $M_{in} = 10^{17}$~GeV
(upper) for different choices of $\lambda_4$:
0.3 (upper left), 0.5 (upper right ) and $M_{in} = 2.4 \times 10^{18}$~GeV (lower) 
for $\lambda_4 = $0.2  (lower left), 
0.3  (lower right). 
}
\label{fig:tb55.1}
\end{figure}

We now turn to the analogous $(m_{1/2}, m_0)$ planes for $\tan \beta = 55$ shown in
Fig.~\ref{fig:tb55}. The upper left panel displays the $\nu$CMSSM case, where
we see the rapid-annihilation heavy Higgs funnel extending up to $(m_{1/2}, m_0)
\sim (1500, 1500)$~GeV. As $M_{in}$ is increased, the heavy Higgs funnel descends rapidly 
into the forbidden charged-LSP region. 
This is seen in the remaining panels of Fig.~\ref{fig:tb55}. In the upper right panel for CFSU(5) with a value of 
$M_{in}$ only slightly larger than $\mgut$,
$M_{in} = 1.4 \times 10^{16}$~GeV, and $\lambda_4 = \lambda_5 = 0.1$, we see that
the rapid-annihilation funnel moves to smaller values of $m_0$, whereas
the other constraints are little affected~\footnote{We draw attention to the
appearance in this panel of a secondary strip of
acceptable relic density running roughly parallel to the boundary of the forbidden 
${\tilde \tau_1}$-LSP region, but with values of $m_0$ about 200~GeV larger.
Here rapid ${\tilde \tau_1} - \overline{{\tilde \tau_1}}$
annihilation via direct-channel $A/H$ poles allows for efficient stau coannihilation with a larger
 $\chi-{\tilde \tau_1}$ mass gap than in the primary stau-coannihilation strip. For the $\tilde \tau_1$
 funnel to be effective at the larger mass gap, one must sit very close to the pole,
 and as a result the secondary strip is very narrow.  Furthermore, the LSP is overdense
 between this and the primary coannihilation strip.
Traces of this feature are also visible in the upper
left panel of Fig.~\ref{fig:tb55} and the upper panels of Fig.~\ref{fig:tb55.1}.}.
Continuing to the case $M_{in} = 10^{17}$~GeV
(lower left), we see that the rapid-annihilation strip has collapsed into a
coannihilation strip along the boundary of the ${\tilde \tau_1}$ LSP region.
Increasing $M_{in}$ further to $2.4 \times 10^{18}$~GeV (lower right), in addition to
this coannihilation strip we see also a light-Higgs rapid-annihilation funnel at
$m_{1/2} \sim 150$~GeV, part of which with $m_0 > 1700$~GeV is compatible with the
$b \to s \gamma$ and LEP chargino and Higgs constraints.

Fig.~\ref{fig:tb55.1} explores the implications of varying $\lambda_4$, keeping $\tan \beta = 55$ fixed.
In the upper left panel, for $\lambda_4 = 0.3, \lambda_5 =0.1$ and $M_{in} = 10^{17}$~GeV, 
comparing with the lower left panel of Fig.~\ref{fig:tb55}, we first notice at small $m_{1/2}$
and large $m_0$ a triangular orange region where there is no consistent electroweak vacuum.
At somewhat larger values of $m_{1/2}$ (lower values of $m_0$) we see a rapid-annihilation funnel
that appears at much larger $m_0$ than the funnels in the upper panels of Fig.~\ref{fig:tb55}.
This reappearance of the rapid-annihilation funnel is due to heavy Higgs bosons becoming 
lighter for larger $\lambda_4$ as 
discussed in detail in Section~\ref{sec:RGEs}. 
When $\lambda_4$ is increased to 0.5 (upper left panel) the electroweak vacuum constraint
advances to larger $m_{1/2}$ (lower $m_0$), and the rapid-annihilation funnel retreats
towards the ${\tilde \tau_1}$ LSP boundary. Qualitatively similar effects are seen in the
lower panels of Fig.~\ref{fig:tb55.1}, for $M_{in} = 2.4 \times 10^{18}$~GeV. In the lower left
panel for $\lambda_4 = 0.2$, we see WMAP-compatible strips in the coannihilation region
near the ${\tilde \tau_1}$ LSP boundary and in the focus-point region near the electroweak
vacuum boundary. When $\lambda_4$ is increased to 0.3 (lower right), a rapid-annihilation
funnel detaches itself from the electroweak vacuum boundary, and moves towards the
${\tilde \tau_1}$ LSP boundary.

It is important to note that although the no-EWSB regions (shaded orange) appear in Figs.~\ref{fig:tb55} and \ref{fig:tb55.1}, they
are of a different nature. In the $\nu$CMSSM case shown in Fig.~\ref{fig:tb55}a (as well as in the CMSSM), the no-EWSB region
appears because $\mu^2 <0$ at large $m_0$ values~\cite{fp}. In CFSU(5) the no-EWSB region appears for the same reason only for small values
of $\lambda_4(\mgut)$, although it is pushed to higher $m_0$ values due to the extra RGE running.
Increasing of $\lambda_4(\mgut)$ leads to a smaller weak scale value of $m^2_{H_d}$ (see discussion for
Fig.~\ref{fig:renn}). At some point $m^2_{H_d}(M_{weak})$ becomes so small that positive $\mu^2$ combined with 
$m^2_{H_d}(M_{weak})$ can no longer compensate the negative $m^2_{H_u}(M_{weak})$, as can be seen from Eq.~\ref{eq:REWSB}, 
yielding $m^2_A <0$ and signals the absence of a consistent electroweak vacuum.
As a consequence, there is no focus-point region in the CFSU(5) that appears close to the $\mu^2 <0$ boundary: at small
$\lambda_4(\mgut)$ it is pushed to very large $m_0$ values and at larger $\lambda_4(\mgut)$ it is preceded by the $m^2_A <0$ region.
The narrow relic density-allowed region visible in the bottom left frame of Fig.~\ref{fig:tb55.1} next to the $m^2_A <0$ boundary is not the
focus point, but rather the rapid-annihilation funnel that will detach from the boundary for larger $\lambda_4$ values.

\begin{figure}
\epsfig{file=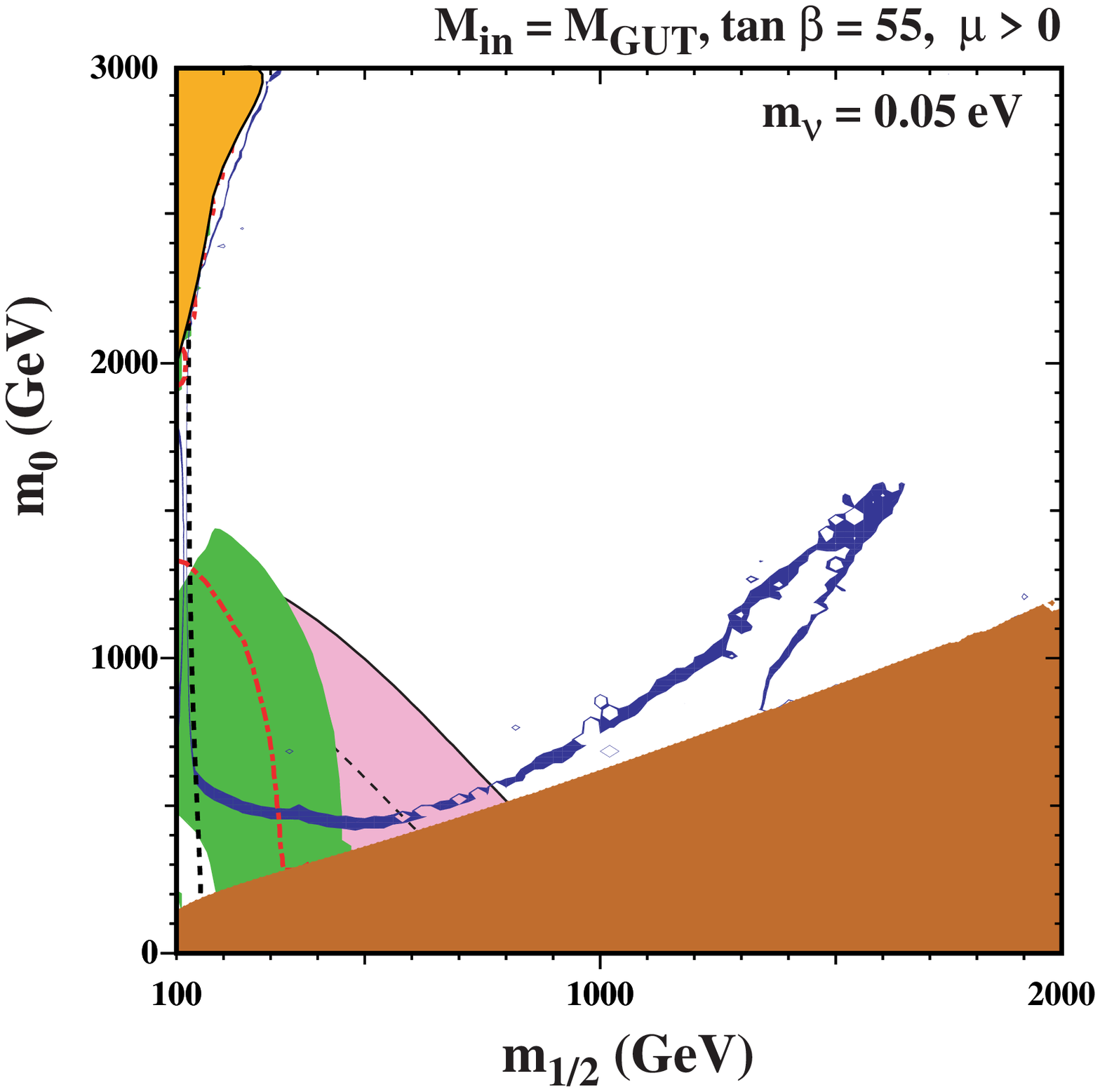,height=8.0cm}
\epsfig{file=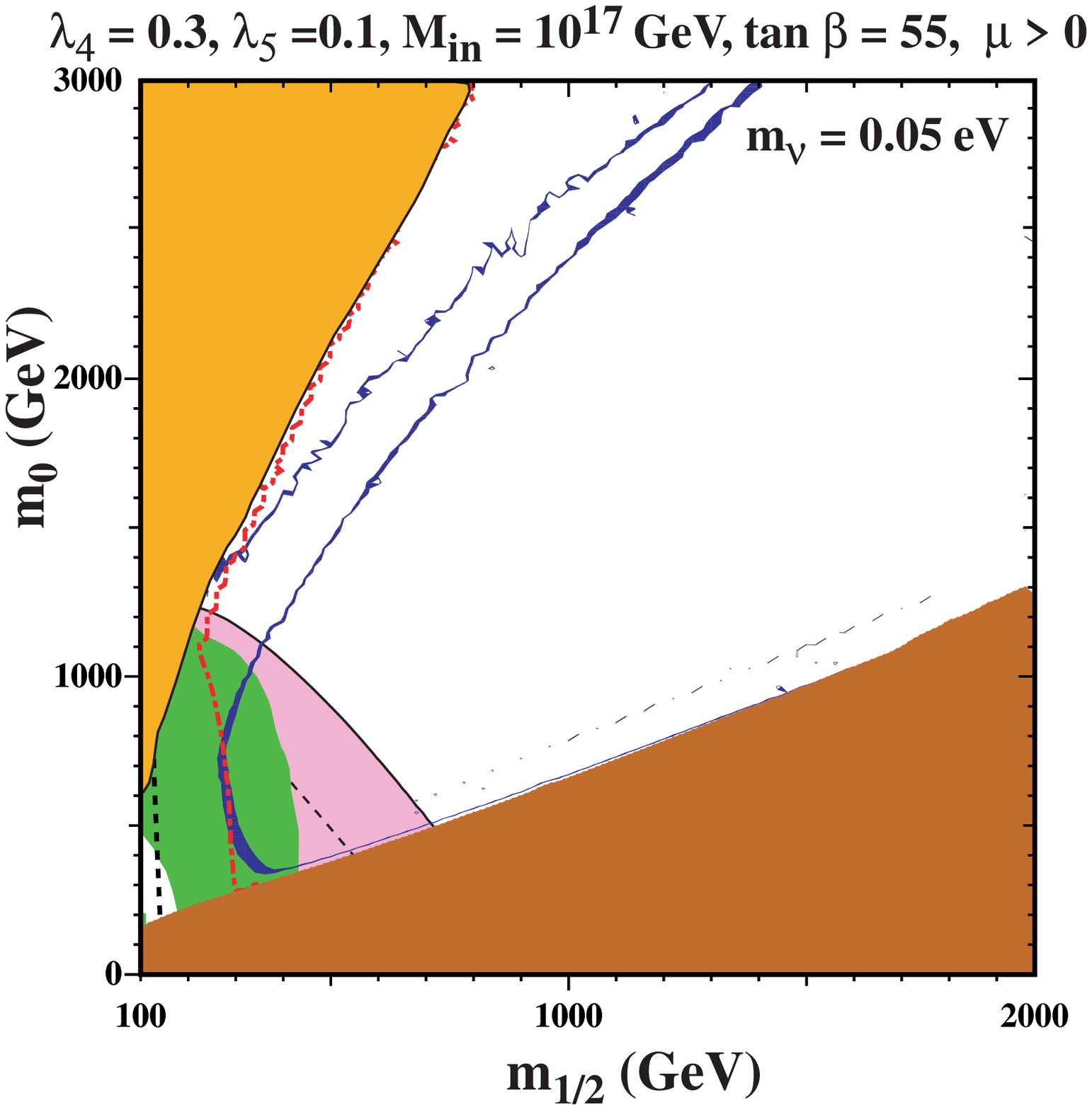,height=8.0cm}
\caption{\it
As for Fig.~\protect\ref{fig:tb10}, for $\tan \beta = 55$ and $m_{\nu_3} = 0.05$ eV, 
with $M_{in} = \mgut$ (left, to be compared with the upper right panel of Fig.~\protect\ref{fig:tb55}) 
and $M_{in} = 10^{17}$ GeV with $\lambda_4$ = 0.3, $\lambda_5$ = 0.1 (right, to be compared with the 
upper left panel of Fig.~\protect\ref{fig:tb55.1}). 
}
\label{varymnu}
\end{figure}

Finally, we show in Fig.~\ref{varymnu}, the effect of lowering our input 
third-generation neutrino mass.  Up to now, we had fixed $m_{\nu_3} = 0.3$~eV
to accentuate the effect of the large neutrino coupling on the running of the RGEs.
A more natural choice which does not require a light neutrino mass degeneracy, in view of cosmological and neutrino oscillation data, might be $m_{\nu_3}$ = 0.05~eV.
We have verified that changing to this choice has negligible effects in almost all cases considered here.
For example, when $\tan \beta = 10$, there would be no visible change in Fig.~\ref{fig:tb10},
as the coannihilation region is known to be very insensitive to the choice of neutrino mass~\cite{msugrarhn} 
when a seesaw neutrino sector is added to the CMSSM.  
The effect of decreasing $m_{\nu_3}$ is only slightly noticeable even when $\tan \beta = 55$ and 
$\lambda_4 = \lambda_5 = 0.1$, as shown
in the left panel of Fig.~\ref{varymnu} where we display the case $M_{in} = \mgut$ and $m_{\nu_3}$ = 0.05~eV. In this case, we have  $M_{N_3} = 1.6 \times 10^{14}$~GeV. 
Comparing with the upper right panel of Fig.~\ref{fig:tb55}, we see that the funnel moves up slightly in 
$m_{0}$, and the focus-point region at large $m_0$ is now more visible in the upper left corner of the figure.
When $M_{in} = 10^{17}$~GeV (not shown), the focus-point region is barely present in the upper left corner 
at the same low values of $m_{1/2}$ and high $m_0$, 
and as expected there is no change in the coannihilation region.
When $\lambda_4$ is increased to 0.3, as in the upper left panel of Fig.~\ref{fig:tb55.1}, 
for $m_{\nu_3}$ = 0.05~eV we find that the funnel region is shifted down slightly to lower $m_0$, 
as shown in the right panel of Fig.~\ref{varymnu}. 
Changes in the remaining panels of  Fig.~\ref{fig:tb55.1} are considerably less pronounced when $m_{\nu_3}$ is lowered to 0.05 eV.

\section{Summary}
\label{sec:concl}

We have studied the parameter space of the minimal flipped SU(5) model with
SSB parameters constrained to be universal at some scale $M_{in} > \mgut$ (CFSU(5)).
We have explored the $M_{in}$ dependences of the ratios between the mass of the LSP $\chi$,
the mass of the lighter stau ${\tilde \tau_1}$ and the heavier MSSM Higgs bosons $A, H$.
As we illustrate by several examples for $\tan \beta = 10$ and 55, these $M_{in}$
dependences affect the locations and even existence of coannihilation strips and
rapid-annihilation funnels, thereby altering the regions of $m_{1/2}$ and $m_0$ in
which the relic $\chi$ density falls within the range favoured by WMAP and other
astrophysical and cosmological data. On the other hand, the WMAP regions are not 
very sensitive to the choice of neutrino mass.

These results reinforce the point already made in~\cite{EMO}, namely that the parts of the
$(m_{1/2}, m_0)$ planes favoured in the CMSSM are substantially modified in scenarios
where $M_{in} > \mgut$. Moreover, as could be expected, these modifications depend
on the GUT model used, being significantly different in CFSU(5) and minimal conventional
SU(5), for example. If supersymmetry is discovered at the LHC and/or in searches for
astrophysical dark matter, these differences may serve as diagnostic tools able to
discriminate between different scenarios for physics at the GUT scale and beyond.

\section*{Acknowledgements}

The work of A.M. and K.A.O. was supported in part by DOE grant DE--FG02--94ER--40823 at the University of Minnesota. 
We thank I.~Gogoladze and Q.~Shafi for many useful discussions. 
K.A.O. also thanks SLAC (supported by the DOE under contract number DE-AC02-76SF00515) and the 
Stanford Institute for Theoretical Physics for their hospitality and support while this work was being finished.


\begin{thebibliography}{99}

\bibitem{EN}
  J.~R.~Ellis and D.~V.~Nanopoulos,
  Phys.\ Lett.\  B {\bf 110}, 44 (1982).

\bibitem{BG}
R.~Barbieri and R.~Gatto,
  Phys.\ Lett.\  B {\bf 110}, 211 (1982).
    
\bibitem{funnel}
  M.~Drees and M.~M.~Nojiri,
  Phys.\ Rev.\  D {\bf 47}, 376 (1993)[arXiv:hep-ph/9207234]; 
  H.~Baer and M.~Brhlik,
  Phys.\ Rev.\  D {\bf 53}, 597 (1996)  [arXiv:hep-ph/9508321] and 
  Phys.\ Rev.\  D {\bf 57}, 567 (1998)  [arXiv:hep-ph/9706509]; 
  H.~Baer, M.~Brhlik, M.~A.~Diaz, J.~Ferrandis, P.~Mercadante, P.~Quintana and X.~Tata,
  Phys.\ Rev.\  D {\bf 63} (2001) 015007
  [arXiv:hep-ph/0005027];
 A.~B.~Lahanas, D.~V.~Nanopoulos and V.~C.~Spanos,
  Mod.\ Phys.\ Lett.\  A {\bf 16} (2001) 1229
  [arXiv:hep-ph/0009065];
  A.~B.~Lahanas and V.~C.~Spanos,
  Eur.\ Phys.\ J.\ C {\bf 23} (2002) 185
  [arXiv:hep-ph/0106345].

\bibitem{cmssm}
J.~R.~Ellis, T.~Falk, K.~A.~Olive and M.~Schmitt,
Phys.\ Lett.\ B {\bf 388} (1996) 97
[arXiv:hep-ph/9607292];
Phys.\ Lett.\ B {\bf 413} (1997) 355
[arXiv:hep-ph/9705444];
J.~R.~Ellis, T.~Falk, G.~Ganis, K.~A.~Olive and M.~Schmitt,
Phys.\ Rev.\ D {\bf 58} (1998) 095002
[arXiv:hep-ph/9801445];
V.~D.~Barger and C.~Kao,
Phys.\ Rev.\ D {\bf 57} (1998) 3131
[arXiv:hep-ph/9704403].
J.~R.~Ellis, T.~Falk, G.~Ganis and K.~A.~Olive,
Phys.\ Rev.\ D {\bf 62} (2000) 075010
[arXiv:hep-ph/0004169].

\bibitem{efgosi}
J.~R.~Ellis, T.~Falk, G.~Ganis, K.~A.~Olive and M.~Srednicki,
Phys.\ Lett.\ B {\bf 510} (2001) 236
[arXiv:hep-ph/0102098].

\bibitem{cmssmnew}
V.~D.~Barger and C.~Kao,
Phys.\ Lett.\ B {\bf 518} (2001) 117
[arXiv:hep-ph/0106189];
L.~Roszkowski, R.~Ruiz de Austri and T.~Nihei,
JHEP {\bf 0108} (2001) 024
[arXiv:hep-ph/0106334];
A.~Djouadi, M.~Drees and J.~L.~Kneur,
JHEP {\bf 0108} (2001) 055
[arXiv:hep-ph/0107316];
U.~Chattopadhyay, A.~Corsetti and P.~Nath,
Phys.\ Rev.\ D {\bf 66} (2002) 035003
[arXiv:hep-ph/0201001];
J.~R.~Ellis, K.~A.~Olive and Y.~Santoso,
New Jour.\ Phys.\  {\bf 4} (2002) 32
[arXiv:hep-ph/0202110];
H.~Baer, C.~Balazs, A.~Belyaev, J.~K.~Mizukoshi, X.~Tata and Y.~Wang,
JHEP {\bf 0207} (2002) 050
[arXiv:hep-ph/0205325];
  R.~L.~Arnowitt and B.~Dutta,
  arXiv:hep-ph/0211417.

\bibitem{cmssmmap}
J.~R.~Ellis, K.~A.~Olive, Y.~Santoso and V.~C.~Spanos,
Phys.\ Lett.\ B {\bf 565} (2003) 176
[arXiv:hep-ph/0303043];
H.~Baer and C.~Balazs,
  JCAP {\bf 0305}, 006 (2003)
  [arXiv:hep-ph/0303114];
A.~B.~Lahanas and D.~V.~Nanopoulos,
  Phys.\ Lett.\  B {\bf 568}, 55 (2003)
  [arXiv:hep-ph/0303130];
U.~Chattopadhyay, A.~Corsetti and P.~Nath,
  Phys.\ Rev.\  D {\bf 68}, 035005 (2003)
  [arXiv:hep-ph/0303201];
   C.~Munoz,
  Int.\ J.\ Mod.\ Phys.\  A {\bf 19}, 3093 (2004)
  [arXiv:hep-ph/0309346].
  
\bibitem{EMO}
  J.~Ellis, A.~Mustafayev and K.~A.~Olive,
  Eur.\ Phys.\ J.\  C {\bf 69}, 201 (2010)
  [arXiv:1003.3677 [hep-ph]].
  
\bibitem{Nanopoulos}
 J.~Jiang, T.~Li, D.~V.~Nanopoulos and D.~Xie,
  Phys.\ Lett.\  B {\bf 677} (2009) 322;
  Nucl.\ Phys.\  B {\bf 830} (2010) 195
  [arXiv:0905.3394 [hep-th]].
T.~Li, J.~A.~Maxin, D.~V.~Nanopoulos and J.~W.~Walker,
  arXiv:1003.4186 [hep-ph];
  arXiv:1007.5100 [hep-ph];
  arXiv:1009.2981 [hep-ph].

  
\bibitem{Antoniadis:1989zy}
  I.~Antoniadis, J.~R.~Ellis, J.~S.~Hagelin and D.~V.~Nanopoulos,
  Phys.\ Lett.\  B {\bf 231}, 65 (1989).
  
\bibitem{Ellis:1992nq}
  J.~R.~Ellis, D.~V.~Nanopoulos and K.~A.~Olive,
  Phys.\ Lett.\  B {\bf 300}, 121 (1993)
  [arXiv:hep-ph/9211325].

\bibitem{Leontaris:1992wp}
S.~A.~Abel,
  Phys.\ Lett.\  B {\bf 234}, 113 (1990);
  G.~K.~Leontaris and J.~D.~Vergados,
  Phys.\ Lett.\  B {\bf 305}, 242 (1993)
  [arXiv:hep-ph/9301291].

\bibitem{infl}
  B.~Kyae, Q.~Shafi,
  Phys.\ Lett.\  {\bf B635}, 247-252 (2006).
  [hep-ph/0510105]; 
    M.~U.~Rehman, Q.~Shafi, J.~R.~Wickman,
  Phys.\ Lett.\  {\bf B688}, 75-81 (2010).
  [arXiv:0912.4737 [hep-ph]].
  
\bibitem{WMAP} 
  E.~Komatsu {\it et al.}  [WMAP Collaboration],
  Astrophys.\ J.\ Suppl.\  {\bf 192}, 18 (2011)
  [arXiv:1001.4538 [astro-ph.CO]].

\bibitem{fsu5}
S.~M.~Barr,
  Phys.\ Lett.\  B {\bf 112}, 219 (1982); 
  S.~M.~Barr,
  Phys.\ Rev.\  D {\bf 40}, 2457 (1989); 
  J.~P.~Derendinger, J.~E.~Kim and D.~V.~Nanopoulos,
  Phys.\ Lett.\  B {\bf 139}, 170 (1984); 
  Q.~Shafi and Z.~Tavartkiladze,
  Phys.\ Lett.\  B {\bf 448}, 46 (1999)
  [arXiv:hep-ph/9811463].

\bibitem{Antoniadis:1987dx}
  I.~Antoniadis, J.~R.~Ellis, J.~S.~Hagelin and D.~V.~Nanopoulos,
  Phys.\ Lett.\  B {\bf 194}, 231 (1987).

\bibitem{mu2}
  J.~L.~Lopez and D.~V.~Nanopoulos,
  Phys.\ Lett.\  B {\bf 251}, 73 (1990); 
  J.~L.~Lopez and D.~V.~Nanopoulos,
  Phys.\ Lett.\  B {\bf 256}, 150 (1991); 
  J.~L.~Lopez and D.~V.~Nanopoulos,
  Phys.\ Lett.\  B {\bf 268}, 359 (1991).
  J.~E.~Kim and H.~P.~Nilles,
  Phys.\ Lett.\  B {\bf 138}, 150 (1984); 
  E.~J.~Chun, J.~E.~Kim and H.~P.~Nilles,
  Nucl.\ Phys.\  B {\bf 370}, 105 (1992);
  J.~A.~Casas and C.~Mu\~noz,
  Phys.\ Lett.\  B {\bf 306}, 288 (1993)
  [arXiv:hep-ph/9302227].

\bibitem{mu3}
  G.~F.~Giudice and A.~Masiero,
  Phys.\ Lett.\  B {\bf 206}, 480 (1988).

\bibitem{dseesaw}
  R.~N.~Mohapatra,
  Phys.\ Rev.\ Lett.\  {\bf 56}, 561 (1986).

\bibitem{Ciuchini:2007ha}
  M.~Ciuchini, A.~Masiero, P.~Paradisi, L.~Silvestrini, S.~K.~Vempati and O.~Vives,
  Nucl.\ Phys.\  B {\bf 783}, 112 (2007)
  [arXiv:hep-ph/0702144].

\bibitem{mt}
 Tevatron Electroweak Working Group and the CDF and D0
  Collaborations,
  arXiv:1007.3178 [hep-ex].

\bibitem{rpp} 
   K.~Nakamura {\it et al.}  [Particle Data Group],
  J.\ Phys.\ G {\bf 37}, 075021 (2010).
  
\bibitem{newBNL} 
 [The Muon g-2 Collaboration],
 {\it Phys. Rev. Lett.} {\bf 92} (2004) 161802, 
 [arXiv:hep-ex/0401008];
 G.~Bennett et al.\ [The Muon g-2 Collaboration],
  {\em Phys.\ Rev.} {\bf D 73} (2006) 072003
  [arXiv:hep-ex/0602035].

\bibitem{g-2}
 M.~Knecht,
  Lect.\ Notes Phys.\  {\bf 629}, 37 (2004)
  [arXiv:hep-ph/0307239];
J.~F.~de Troconiz and F.~J.~Yndurain,
  Phys.\ Rev.\  D {\bf 71}, 073008 (2005)
  [arXiv:hep-ph/0402285];
   M.~Passera,
  arXiv:hep-ph/0411168;
      D.~Stockinger,
  J.\ Phys.\ G {\bf 34} (2007) R45
  [arXiv:hep-ph/0609168];
     J.~Miller, E.~de~Rafael and B.~Roberts,
   {\em Rept.\ Prog.\ Phys.} {\bf 70} (2007) 795
   [arXiv:hep-ph/0703049];
 J.~Prades, E.~de Rafael and A.~Vainshtein,
  arXiv:0901.0306 [hep-ph];
    F.~Jegerlehner and A.~Nyffeler,
  Phys.\ Rept.\  {\bf 477}, 1 (2009)
  [arXiv:0902.3360 [hep-ph]];
  J.~Prades,
  Acta Phys.\ Polon.\ Supp.\  {\bf 3}, 75 (2010)
  [arXiv:0909.2546 [hep-ph]];
    T.~Teubner, K.~Hagiwara, R.~Liao, A.~D.~Martin and D.~Nomura,
  arXiv:1001.5401 [hep-ph];
M.~Davier, A.~Hoecker, B.~Malaescu and Z.~Zhang,
  arXiv:1010.4180 [hep-ph].

\bibitem{bsgex}
S.~Chen {\it et al.}  [CLEO Collaboration],
Phys.\ Rev.\ Lett.\  {\bf 87} (2001) 251807
[arXiv:hep-ex/0108032];
P.~Koppenburg {\it et al.}  [Belle Collaboration],
Phys.\ Rev.\ Lett.\  {\bf 93} (2004) 061803
[arXiv:hep-ex/0403004].
B.~Aubert {\it et al.}  [BaBar Collaboration],
arXiv:hep-ex/0207076;
E.~Barberio {\it et al.}  [Heavy Flavor Averaging Group (HFAG)],
  arXiv:hep-ex/0603003.

\bibitem{ssard} Information about this code is available from K.~A.~Olive: it contains important contributions 
from T.~Falk, G.~Ganis, A.~Mustafayev, J.~McDonald, K.~A.~Olive, P.~Sandick, Y.~Santoso and M.~Srednicki. 


\bibitem{FeynHiggs}
S.~Heinemeyer, W.~Hollik and G.~Weiglein,
{\it Comput.\ Phys.\ Commun.\ } {\bf 124} (2000) 76 
[arXiv:hep-ph/9812320];
S.~Heinemeyer, W.~Hollik and G.~Weiglein,
{\it Eur.\ Phys.\ J.\ C} {\bf 9} (1999) 343 
[arXiv:hep-ph/9812472];
  M.~Frank, T.~Hahn, S.~Heinemeyer, W.~Hollik, H.~Rzehak and G.~Weiglein,
  JHEP {\bf 0702} (2007) 047
  [arXiv:hep-ph/0611326].

\bibitem{isajet} 
  ISAJET, by H.~Baer, F.~Paige, S.~Protopopescu and X.~Tata, arXiv:hep-ph/0312045.
  
\bibitem{Battaglia:2001zp}
  M.~Battaglia {\it et al.},
  Eur.\ Phys.\ J.\  C {\bf 22}, 535 (2001)
  [arXiv:hep-ph/0106204];
 M.~Battaglia, A.~De Roeck, J.~R.~Ellis, F.~Gianotti, K.~A.~Olive and L.~Pape,
  Eur.\ Phys.\ J.\  C {\bf 33}, 273 (2004)
  [arXiv:hep-ph/0306219];
 A.~De Roeck, J.~R.~Ellis, F.~Gianotti, F.~Moortgat, K.~A.~Olive and L.~Pape,
  Eur.\ Phys.\ J.\  C {\bf 49}, 1041 (2007)
  [arXiv:hep-ph/0508198].
  
\bibitem{MSSMRGEs}
  N.~K.~Falck,
  Z.\ Phys.\  C {\bf 30}, 247 (1986); 
S.~P.~Martin and M.~T.~Vaughn,
Phys.\ Rev.\ D {\bf 50} (1994) 2282
[arXiv:hep-ph/9311340].

\bibitem{msugrarhn} 
  H.~Baer, M.~A.~Diaz, P.~Quintana and X.~Tata,
  JHEP {\bf 0004}, 016 (2000)
  [arXiv:hep-ph/0002245]; 
  S.~T.~Petcov, S.~Profumo, Y.~Takanishi, C.~E.~Yaguna,
  Nucl.\ Phys.\  {\bf B676}, 453-480 (2004).
  [hep-ph/0306195];
  L.~Calibbi, Y.~Mambrini and S.~K.~Vempati,
  JHEP {\bf 0709}, 081 (2007)
  [arXiv:0704.3518 [hep-ph]];
  V.~Barger, D.~Marfatia and A.~Mustafayev,
  Phys.\ Lett.\  B {\bf 665}, 242 (2008)
  [arXiv:0804.3601 [hep-ph]]; 
   M.~E.~Gomez, S.~Lola, P.~Naranjo and J.~Rodriguez-Quintero,
  JHEP {\bf 0904}, 043 (2009)
  [arXiv:0901.4013 [hep-ph]];
  K.~Kadota, K.~A.~Olive and L.~Velasco-Sevilla,
  Phys.\ Rev.\  D {\bf 79}, 055018 (2009)
  [arXiv:0902.2510 [hep-ph]];
  V.~Barger, D.~Marfatia, A.~Mustafayev and A.~Soleimani,
  Phys.\ Rev.\  D {\bf 80}, 076004 (2009)
  [arXiv:0908.0941 [hep-ph]];
  K.~Kadota and K.~A.~Olive,
  Phys.\ Rev.\  D {\bf 80}, 095015 (2009)
  [arXiv:0909.3075 [hep-ph]];
 M.~Hirsch, L.~Reichert, W.~Porod,
  [arXiv:1101.2140 [hep-ph]].

\bibitem{stauco}
J.~R.~Ellis, T.~Falk and K.~A.~Olive,
Phys.\ Lett.\ B {\bf 444} (1998) 367
[arXiv:hep-ph/9810360];
J.~R.~Ellis, T.~Falk, K.~A.~Olive and M.~Srednicki,
Astropart.\ Phys.\  {\bf 13} (2000) 181
[Erratum-ibid.\  {\bf 15} (2001) 413]
[arXiv:hep-ph/9905481];
R.~Arnowitt, B.~Dutta and Y.~Santoso,
Nucl.\ Phys.\ B {\bf 606} (2001) 59
[arXiv:hep-ph/0102181];
M.~E.~G\'omez, G.~Lazarides and C.~Pallis,
Phys. Rev. D {\bf D61} (2000) 123512
[arXiv:hep-ph/9907261];
  Phys.\ Lett. {\bf B487} (2000) 313
[arXiv:hep-ph/0004028];
  Nucl. Phys. B {\bf B638} (2002) 165
[arXiv:hep-ph/0203131];
T.~Nihei, L.~Roszkowski and R.~Ruiz de Austri,
  JHEP {\bf 0207} (2002) 024
[arXiv:hep-ph/0206266].

\bibitem{flippedDM}
  J.~R.~Ellis, J.~S.~Hagelin, S.~Kelley, D.~V.~Nanopoulos, K.~A.~Olive,
  Phys.\ Lett.\  {\bf B209}, 283 (1988); 
  M.~Drees, X.~Tata,
  Phys.\ Lett.\  {\bf B206}, 259 (1988); 
  J.~R.~Ellis, J.~S.~Hagelin, S.~Kelley, D.~V.~Nanopoulos,
  Nucl.\ Phys.\  {\bf B311}, 1 (1988);
   J.~McDonald,
  Phys.\ Lett.\  B {\bf 225}, 133 (1989);
  I.~Gogoladze, R.~Khalid, S.~Raza and Q.~Shafi,
  Mod.\ Phys.\ Lett.\  A {\bf 25}, 3371 (2010)
  [arXiv:0912.5411 [hep-ph]].

\bibitem{Griest:1990kh}
  K.~Griest and D.~Seckel,
  Phys.\ Rev.\  D {\bf 43}, 3191 (1991).

\bibitem{hfag}
E.~Barberio {\it et al.}  [Heavy Flavor Averaging Group (HFAG)],
  arXiv:hep-ex/0603003.
  
\bibitem{fp}
  J.~L.~Feng, K.~T.~Matchev and T.~Moroi,
  Phys.\ Rev.\ Lett.\  {\bf 84}, 2322 (2000)
  [arXiv:hep-ph/9908309],
  and  
  Phys.\ Rev.\  D {\bf 61}, 075005 (2000)
  [arXiv:hep-ph/9909334]; 
  J.~L.~Feng, K.~T.~Matchev and F.~Wilczek,
  Phys.\ Lett.\  B {\bf 482}, 388 (2000)
  [arXiv:hep-ph/0004043].

\bibitem{bsgprocedure}
  J.~R.~Ellis, S.~Heinemeyer, K.~A.~Olive, A.~M.~Weber and G.~Weiglein,
  JHEP {\bf 0708}, 083 (2007)
  [arXiv:0706.0652 [hep-ph]].

\bibitem{gam}
  G.~Degrassi, P.~Gambino and G.~F.~Giudice,
  JHEP {\bf 0012}, 009 (2000)
  [arXiv:hep-ph/0009337].
 
\bibitem{LEPsusy}
Joint LEP~2 Supersymmetry Working Group,
{\it Combined LEP Chargino Results up to 208 GeV}, \\
{\tt http://lepsusy.web.cern.ch/lepsusy/www/inos{\_}moriond01/%
charginos{\_}pub.html}.
  
\bibitem{LEPHiggs}
LEP Higgs Working Group for Higgs boson searches, OPAL Collaboration,
ALEPH Collaboration, DELPHI Collaboration and L3
Collaboration,
Phys.\ Lett.\ B {\bf 565} (2003) 61 [arXiv:hep-ex/0306033].
{\it Search for neutral Higgs bosons at LEP}, paper submitted to 
ICHEP04, Beijing,
LHWG-NOTE-2004-01, ALEPH-2004-008, DELPHI-2004-042, L3-NOTE-2820,
OPAL-TN-744, \\
{\tt http://lephiggs.web.cern.ch/LEPHIGGS/papers/August2004{\_}MSSM/index.html}.


\bibitem{hfunnel} 
  P.~Nath and R.~L.~Arnowitt,
  Phys.\ Rev.\ Lett.\  {\bf 70}, 3696 (1993)
  [arXiv:hep-ph/9302318]; 
  A.~Djouadi, M.~Drees and J.~L.~Kneur,
  Phys.\ Lett.\  B {\bf 624}, 60 (2005)
  [arXiv:hep-ph/0504090].

\end{thebibliography}
\end{document}